\documentclass[prd,twocolumn,showpacs,amsmath,amssymb,nofootinbib]{revtex4}

\newcommand\lsim{\lesssim}
\newcommand\gsim{\gtrsim}

\newcommand\vev[1]{{\langle {#1} \rangle}}
\renewcommand\({\left(}
\renewcommand\){\right)}
\renewcommand\[{\left[}
\renewcommand\]{\right]}

\newcommand\eq[1]{Eq.~(\ref{#1})}
\newcommand\eqs[2]{Eqs.~(\ref{#1}) and (\ref{#2})}
\newcommand\eqss[3]{Eqs.~(\ref{#1}), (\ref{#2}), and (\ref{#3})}

\newcommand\eqst[2]{Eqs.~(\ref{#1})--(\ref{#2})}

\newcommand\eqreff[1]{(\ref{#1})}
\newcommand\eqsref[2]{(\ref{#1}) and (\ref{#2})}

\newcommand\pa{\partial}

\newcommand\ee{\end{equation}}
\newcommand\be{\begin{equation}}
\newcommand\eea{\end{eqnarray}}
\newcommand\bea{\begin{eqnarray}}

\newcommand\mpl{M_{\rm P}}

\newcommand{\dlabel}[1]{\label{#1} \ \ \ \ \ \ \ \ #1\ \ \ \ \ \ \ \ }

\def\calp{{\cal P}}

\def\calv{{\cal V}}

\newcommand\bfk{{\mathbf k}}

\newcommand\bfp{{\mathbf p}}
\newcommand\bfq{{\mathbf q}}

\newcommand\bfx{{\mathbf x}}

\newcommand\GeV{\,\mbox{GeV}}
\newcommand\MeV{\,\mbox{MeV}}

\newcommand\Mpc{\,\mbox{Mpc}}

\newcommand\msun{M_\odot}

\newcommand\sub[1]{_{\rm #1}}
\newcommand\su[1]{^{\rm #1}}

\newcommand\mone{^{-1}}
\newcommand\mtwo{^{-2}}
\newcommand\mthree{^{-3}}

\newcommand\mhalf{^{-1/2}}
\newcommand\half{^{1/2}}

\newcommand\quarter{^{1/4}}

\newcommand{\fnl}{{f\sub{NL}}}

\newcommand{\tnl}{{\tau\sub{NL}}}

\newcommand{\ma}{\sub{max}}
\newcommand{\mi}{\sub{min}}

\newcommand\kmax{{k\sub{max}}}
\newcommand\bfkp{{{\bfk}'}}

\newcommand\tpq{{(2\pi)^3}}

\newcommand{\zetag}{\zeta\sub{inf}}
\newcommand\sigmas{\sigma^2}
\newcommand\sigmae{\sigma\sub e}

\newcommand\lmone{L^{-1}}
\newcommand\zetasig{\zeta_\sigma}
\newcommand\zetas{\zeta_\sigma}
\newcommand\ts{t_\sigma}

\newcommand\calps{\calp_\sigma}

\newcommand\calpz{\calp_\zeta}

\newcommand\rhoin{\rho\sub{inf}}
\newcommand\rhos{\rho_\sigma}
\newcommand\oms{\Omega_\sigma}

\newcommand\rng{r\sub{ng}}
\newcommand\os{\sub{os}}
\newcommand\so{\sigma\os}

\newcommand\dss{\delta\sigma\sub e}

\newcommand\smax{\sigma\sub{max}}

\begin{document}

\title{Non-gaussianity and cosmic uncertainty in curvaton-type models}

\author{David H.~Lyth}

\affiliation{{\it Physics Department, Lancaster University, Lancaster
LA1 4YB, UK}}

\begin{abstract}
In curvaton-type models, observable 
non-gaussianity of the curvature perturbation would come 
from a 
contribution  of the form $(\delta\sigma)^2$, where $\delta\sigma$ is 
gaussian.  
I analyse this situation 
allowing  $\delta\sigma$ to be scale-dependent. The actual curvaton model is
considered in more detail than before, including its cosmic uncertainty and
anthropic status. The status of curvaton-type models after WMAP year three
data is considered.
\end{abstract}
%\pacs{98.80.Cq}
\maketitle

\section{Introduction} 

It is generally agreed that the primordial curvature perturbation
$\zeta$
 is caused by the perturbation of one or more scalar fields,
those perturbations being generated on  each scale  at horizon exit
during inflation. 
In curvaton-type models, a significant or dominant contribution to $\zeta$
is generated after slow-roll inflation ends, by a field $\sigma$ 
whose potential
is too flat to affect the inflationary dynamics.

The body of this paper is in three sections. Section \ref{s2}
focuses on a 
perturbation of the form
\bea
\zeta(\bfx) &=&  \zetag(\bfx) + \zetasig(\bfx) \nonumber \\
\zetasig(\bfx)  &\equiv&  b\delta\sigma(\bfx) + {\delta\sigma}^2(\bfx) 
\label{withlin} 
.
\eea
(The last term is written with the compact notation $\delta\sigma^2
\equiv (\delta \sigma)^2$, which will be used consistently.)
Following \cite{bl}, the spectrum, bispectrum and trispectrum of the 
perturbation are calculated, allowing for the first time spectral
tilt in the spectrum of $\delta\sigma$. 

In Section \ref{four}, the generation of $\zeta$ is described using the 
$\delta N$ formalism. In curvaton-type models $\zeta$ is given by \eq{withlin}
or its multi-field generalization.
Non-gaussianity and scale-dependence are treated
together, building on the   separate discussions of
 \cite{ss,lr}.

 The prediction
for $\zeta$ has cosmic uncertainty because it
depends on the average value of the curvaton-like field
in our part of the universe.  One may assume that the 
probability distribution for this 
average typically is quite flat up to cutoff. The resulting probability
distribution for $\zeta$ (the `prior' for anthropic considerations)
is model-dependent.  
%This  section also shows (out of line with the paper's title) how to handle
%scale-dependent non-gaussianity  generated during multi-component
%inflation.

Section \ref{five} considers the actual curvaton model. A master formula
for $\zeta$ is presented, including 
all known versions of the model. Assuming that the curvaton 
contribution dominates and that the curvaton has negligible evolution
after inflation, the cosmic uncertainty and anthropic status of the 
curvaton model is described,  extending the recent
 work of  Garriga and Vilenkin
\cite{gv} and 
 Linde and Mukhanov \cite{lm2}.  
Finally, in Section \ref{s6} we look at the status of  curvaton-type,
in the light of the recent measurement of negative spectral tilt for 
the curvature perturbation.

Standard material covered in for instance \cite{treview,book} is
taken for granted throughout, with fuller explanation given for 
more recent developments.

\section{Calculating the correlators}

\label{s2}

 For convenience it is assumed that the spatial average
of  $\delta\sigma$ vanishes.
\be
\overline{\delta\sigma} = 0
\label{vevs}
.
\ee
This  requirement is not essential,\footnote
{In \cite{bl} and elsewhere, a contribution $-\overline{{\delta\sigma}^2}$ 
was added to $\zeta$ to make
 $\overline{\zeta}=0$. That has no effect on the calculations, which deal
only with Fourier modes of $\zeta$ with nonzero wavenumber.}
because if \eq{withlin} were valid with some  $\overline{\delta\sigma}
\neq 0$, one could arrive at 
$\overline{\delta\sigma}=0$ by making  the redefinitions
\bea
\delta\sigma &\to& \delta\sigma - \overline{\delta\sigma}
\nonumber \\
b &\to&  b + 2\overline{\delta\sigma}
\label{beq}
.
\eea

\subsection{Working in a box}
\label{one}

A generic cosmological perturbation, evaluated at some 
instant, will be denoted by by $g(\bfx)$ and  its Fourier components by
\be
g_\bfk =  \int d^3x\, e^{i\bfk\cdot\bfx} \,g(\bfx) 
.
\ee
The integral goes over 
 a box of size $L$, within which the stochastic properties are to be defined.
To describe the cmb anisotropy the box should be 
much bigger than the size $H_0\mone$ of the 
observable Universe.

 Since the box introduces periodic
boundary conditions, one requires that 
physically significant wavelengths are much shorter than the box size,
corresponding to $kL\gg 1$. One can then regard the wave vector $\bfk$
as a continuous variable.

 To describe
the stochastic properties of cosmological perturbations within the box,
one formally invokes an ensemble of universes and takes expectation values
for observable quantities. The zero mode of each perturbation, corresponding
to the spatial average within the box, is not regarded as a stochastic 
variable. The nonzero modes have zero expectation value, $\vev {g_\bfk}=0$.
(Both of these features are predicted by the inflationary cosmology.)
It is  usually supposed that
the observable Universe
corresponds  to a typical member of the ensemble, so that the expectation
values apply. 

Since the stochastic 
properties are supposed to be invariant under translations and rotations
(reflecting, within the inflationary paradigm, the invariance of the vacuum)
a sampling of the ensemble in a given region may be
 regarded as a sampling of different locations for that region.
One can say then, that within the box of size
$L$ we are dealing with the actual Universe, and that the expectation values
refer to the location of the observable Universe within the box.

If  $\ln(LH_0)$ is  not  exponentially large, it should be safe to assume
that our
location within the box is typical.
On the other hand,  the observable Universe may be  part of a very large
 region around us with the same stochastic properties; a region so large
that $\ln(LH_0)$ can be exponentially large. This is what happens
within the inflationary cosmology, if inflation lasted for an exponentially
large number of Hubble times before our Universe left the horizon. 
If such a super-large box is used, one should bear in mind the possibility
that our location is untypical.

The  spectrum $P_g(k)$ is 
 defined by
\be
\vev{g_\bfk\, g_\bfkp}= \tpq \delta^{(3)}(\bfk+\bfkp ) P_g(k)
\label{calpg}
.
\ee
It is useful to define  
 $\calp_g \equiv (k^3/2\pi^2) P_g$,  also called the spectrum.
After smoothing on a scale $R$, the  variance is  
\be
\vev{ g^2(\bfx) } = 
\int^{R\mone}_{\lmone} \frac{dk}{k} \,\calp_g(k)
\label{variance}
.
\ee
The spectral index $n_g$ and the spectral tilt $t_g$ are  defined as
\be
t_g\equiv n_g-1\equiv \frac{d\ln g}{d\ln k}
\label{tiltdef}
.
\ee
For constant tilt, $\calp_g\propto k^{t_g}$.

If $\calp_g(k)$ is sufficiently flat and the range of $k$ is
not too big,
\be
\vev{g^2} \sim \calp_g
\label{specinterp}
.
\ee 
If instead it rises steeply, $\vev{g^2}\sim \calp_g(R\mone)$.
In either case,  the spectrum of a quantity is roughly its mean-square.
This interpretation of the spectrum is implied in many discussions,
including some in the present paper, but it should be applied with
caution.

{}On cosmological scales  $\calp_\zeta$ is almost scale-invariant with
$\calp_\zeta\half =5\times 10^{-5}$. 
At the $2\sigma$ level, the tilt is 
constrained by observation \cite{obs} to 
something like $t_\zeta       = -0.03\pm 0.04$. 

If the two-point correlator is the only (connected) one, the probability
distribution of $g(\bfx)$ is gaussian.
Non-Gaussianity is signaled by additional connected correlators. 
Data are at present consistent with the hypothesis that $\zeta$
is perfectly gaussian, but they  might not be  in the future.

The bispectrum $B_g$ is defined by 
\be \vev{ g_{\bfk_1} g_{\bfk_2}
g_{\bfk_3} } = \tpq \delta^{(3)}(\bfk_1+\bfk_2+\bfk_3)
B_g(k_1,k_2,k_3) 
\label{bg}
.
\ee
Instead of $B_\zeta$ it is more convenient to consider
$\fnl(k_1,k_2,k_3)$, defined by \cite{maldacena}\footnote
{The sign and the prefactor make this definition coincide with the original
one \cite{spergel} in first-order cosmological  perturbation theory,
where $\fnl$ was defined with respect to the Bardeen potential
which was taken to be $\Phi= \frac35 \zeta$.
(In many theoretical works, including previous works by the present author,
$\fnl$ is defined with the opposite sign.)
At second order, which as we see later may be needed if $|\fnl|\lsim 1$,
$\Phi$ and $\zeta$ are completely different functions and
$\fnl$ defined with respect to $\Phi$ has nothing to do with the
$\fnl$ of the present paper.
Unfortunately,
both definitions are in use at the second-order level.}
\be
 B_\zeta(k_1,k_2,k_3)
= \frac65\fnl \[ P_\zeta(k_1)  P_\zeta(k_2) +\,{\rm cyclic} \] 
\label{fnldef} ,
\ee
where the permutations are of $\{k_1,k_2,k_3\}$.

%The bispectrum determines, in particular, the skewness of the probability 
%distribution of $g(\bfx)$. 
%Given the interpretation \eqreff{specinterp}, the amount of non-gaussianity
%generated by the bispectrum will be small if
%\be
%|\fnl| \lsim \calpz\mhalf \sim 10^4
%.
%\ee
Current observation  \cite{komatsu,new} 
 gives  at $2\sigma$ level
\be
- 27 <  \fnl < 121
.
\ee
Absent a detection, observation will eventually \cite{bkmr} 
 bring this down to  $|\fnl|\lsim 1$. 
At that level, the comparison of theory with observation
will require second-order cosmological perturbation theory, whose development
is just beginning \cite{bmr}.

The  trispectrum $T_g$ is   defined in terms of the
connected four-point correlator by
as
\be
\vev{g_{\bfk_1} g_{\bfk_2}g_{\bfk_3}g_{\bfk_4}}_c=
(2\pi)^3\delta^{(3)}(\bfk_1+\bfk_2+\bfk_3+\bfk_4) T_g
\label{tg}
\,.
\ee
It is a function of six scalars,  defining  the
quadrilateral formed by $\{\bfk_1,\bfk_2,\bfk_3,\bfk_4\}$.
It is convenient to consider $\tnl$ defined by \cite{bl}
\be
T_\zeta =\frac12\tau\sub{NL}  P_\zeta(k_1) P_\zeta(k_2) P_\zeta(k_{14})
\,+\,23\,{\rm perms.} 
\,,
\label{taudef}
\ee
 In this expression, 
$\bfk_{ij}\equiv      
\bfk_i+\bfk_j$, and the permutations are of 
$\{\bfk_1,\bfk_2,\bfk_3,\bfk_4 \}$ giving actually 12 distinct terms.

%The trispectrum generates, in particular, the kurtosis of the probability
%distribution of $g(\bfx)$.
%Given the interpretation \eqreff{specinterp}, 
%the amount of  non-gaussianity generated by the trispectrum will be small if
%$\tnl \lsim \calpz\mone \sim 10^9$.
Current observation gives something like \cite{kk}
$\tnl \lsim 10^4$, and absent a detection PLANCK data will give something like
$\tnl \lsim 300$. 

\subsection{The correlators}
\label{onea}

If $\zeta$ is given by \eq{withlin} its spectrum, bispectrum and trispectrum
are given by the following expressions, with all higher connected
correlators vanishing;
\bea
\calp_\zeta &=& \calp_{\zetag} + \calp_{\zetas} \label{speczero} \\
\calp_{\zetas} &=& \calp_{\zetas}\su{linear} + \calp_{\zetas}\su{quad}
 \label{specfirst} \\
\calp_{\zetas}\su{linear} &=& b^2 \calps \\
\calp_{\zetas}\su{quad}&=& 
\calp_{\delta\sigmas} = \frac{k^3}{2\pi} \int_{\lmone} d^3p\frac{
\calps(p)\calps(|\bfp-\bfk|) }{ p^3 |\bfp-\bfk|^3 }
\label{psigs} \\
B_\zeta &=& B_{\zetas}\su{linear} + B_{\zetas}\su{quad}
 \label{bfirst}  \\
B_{\zetas}\su{linear} &=& 8\pi^4 b^2 \( \frac{ \calps(k_1)  \calps(k_2)
}{
k_1^3 k_2^3 }  + {\,\rm cyclic} \) \label{btree} \\
B_{\zetas}\su{quad} &=& B_{\delta\sigmas} \nonumber \\
&=& 
 (2\pi)^3 \int_{\lmone} 
d^3p 
\frac{
\calps(p)\calps(p_1) \calps(p_2)
}{
p^3  p_1^3 p_2^3  }
\label{bsigs} \\
T_\zeta &=& T_{\zetas}\su{linear} + T_{\zetas}\su{quad} \label{tfirst} \\
T_{\zetas}\su{linear} &=& 8\pi^6 b^2 
\frac{  P_\zeta(k_1) P_\zeta(k_2) P_\zeta(k_{14}) }{k_1^3 k_2^3 k_{14}^3 }
\nonumber  \\
&+& \,23\,{\rm perms.}  \label{ttree} \\
T_{\zetas}\su{quad} &=& T_{\delta\sigmas} \nonumber \\
&=&  4\pi^5 \int_{\lmone} d^3p 
\frac{\calps(p)\calps(p_1) \calps(p_2)\calps(p_{24}) }
{p^3 p_1^3 p_2^3 p_{24}^3 } \nonumber \\
&+& \,23\,{\rm perms} \label{tsigs}
.
\eea
The 24 terms in \eq{ttree} are actually 12 pairs of identical terms,
 and the  24 terms in \eq{tsigs} are actually 3 octuplets of identical
terms.
In the integrals  $p_1\equiv |\bfp-\bfk_1|$, $p_2\equiv |\bfp+\bfk_2|$ and
$p_{24}\equiv |\bfp + \bfk_{24}|$. 
The subscript $\lmone$ indicates that the integrand 
is set equal to zero in a sphere of radius 
$\lmone$ around each singularity.
The integral  \eqreff{psigs} was given in  \cite{myaxion}, 
and the integrals  \eqsref{bsigs}{tsigs} were  given in \cite{bl} except that
$\calps$ was taken to be scale-independent. The term $B_{\zetas}\su{linear}$
  was given in \cite{spergel} and the term $T_{\zetas}\su{linear}$
was given in \cite{okamoto,bl}.

In the language of field theory, these expressions are obtained by contracting
pairs of fields, after using the convolution
\be
(\delta\sigmas)_\bfk = \frac1\tpq \int d^3q \delta\sigma_\bfq \delta\sigma_{\bfk-\bfq}
\label{phisk}
\,.
\ee
The terms labeled  `quad' are generated purely by products of 
three $\delta\sigmas$ terms,
while the terms labeled `linear' are generated by a product of two 
$b\delta\sigma$ terms and one $\delta\sigmas$ term.
Their evaluation is best done using Feynman-like graphs (cf.~
\cite{cs,zrl}).  The terms labeled `linear' come
from   tree-level diagrams, while those labeled `quad' come from 
 closely-related one-loop diagrams.

Since $\delta\sigma$ is gaussian, its stochastic properties are determined entirely
by its spectrum $\calps$. The correlators (hence all stochastic 
properties) of ${\delta\sigma}^2$ are
also determined by $\calps$. By examining the large-$p$ behaviour
of  \eqs{psigs}{bsigs}, we see that
{\em the correlators  of ${\delta\sigma}^2$ on a given scale are insensitive
to the spectrum  of $\delta\sigma$ on much smaller scales.}\footnote
{To be precise,
\eqss{psigs}{bsigs}{tsigs}  converge  at $p\gg k$ 
if the tilt $\ts$ is below respectively $3/2$, $2$ and $9/4$.
These conditions are well satisfied 
 within the inflationary  paradigm, which makes  $|\ts|$ well below $1$.}
In contrast with the case of quantum field theory, there is no divergence
in the  ultra-violet (large $k$) regime.

By examining the behaviour of \eqs{psigs}{bsigs} near the singularities,
we see that  with sufficiently large positive tilt, the correlators
 of ${\delta\sigma}^2$ on a given scale are  insensitive to the
spectrum of $\delta\sigma$ also on much bigger scales.
More will be said later about this infra-red regime. 

The  integral  \eqreff{psigs}
can be evaluated exactly \cite{myaxion} to give
\be
\calp_{\delta\sigmas}(k) = 4 \calps^2 \ln(kL) \label{exact}
.
\ee
 The  integral \eqreff{bsigs}
 can be estimated as follows.  Focusing on the singularity
$p=0$, one can consider a sphere around it with radius $k$ a bit less than
$\min\{k_1,k_2\}$. The contribution from this sphere gives
\be
B_{\delta\sigmas} \simeq  \frac{32\pi^4}{k_1^3 k_2^3}
\int^k_{\lmone} \frac {dp}{p} 
=\frac{32\pi^4}{k_1^3 k_2^3} \ln(kL)
.
\ee
A similar sphere around each of the other two 
singularities gives a similar contribution, and the contribution from these
three spheres should be dominant because the integrand at large $p$ goes like
 $p^{-9}$. (Applying this argument to the integral \eqreff{psigs}
happens to give exactly \eq{exact}.)
 Evaluating $\fnl$ we arrive at the estimate
 \cite{bl}
\bea
\frac35 \fnl &=& b^2 \frac{\calps^2}{\calpz^2} + \frac35 
f\sub{NL}\su{quad} \\
\frac35 f\sub{NL}\su{quad}  &\simeq & 
4  \frac{ \calps^3  }{ \calp_\zeta^2 }  \ln(kL) \\
&=& \sqrt \frac 1{2\ln(kL)}  
 \( \frac{\calp_{\delta\sigmas}}{\calp_\zeta} \)^\frac32 \calp_\zeta \mhalf 
\label{fnlflat} 
\eea
with  $k=\min\{k_1,k_2,k_3\}$. A similar estimate for the trispectrum gives
\cite{bl}
\bea
\tnl &=& 4 b^2  \frac{\calps^3}{\calpz^3} + \tau\sub{NL}\su{quad}  \\
\tau\sub{NL}\su{quad}  &\simeq & 16 \frac{\calps^4}{\calp_\zeta^3} \ln(kL) \\
&\simeq &  \frac1{\ln kL} \( \frac{\calp_{\delta\sigmas}}{\calp_\zeta} \)^2 
\frac1{\calp_\zeta}
,
\eea
with $k=\min\{k_i,k_{jm} \}$. 

It is easy to  repeat these estimates for the case of constant nonzero tilt,
$\ts \equiv d\ln \calps/d \ln k$.
To avoid  rather cumbersome expressions,  I give the result for the 
bispectrum only in the regime where  the $k_i$  
have an approximate common value $k$, and the result for the trispectrum
only in the regime where both $k_i$ and $k_{ij}$ have an approximate
common value $k$. Then, the only effect of tilt is to replace $\ln(kL)$ by
%\bea
%\calpss(k) &\simeq&  4 \calp^2_\delta\sigma(k) y(kL)  \\
%\frac35f\sub{NL}^\delta\sigmas &\simeq& 
%\frac 1{\sqrt 2} y\mhalf(kL) \( \frac{\calpss(k)}{\calp_\zeta(k)}
%\)^\frac32 \calp_\zeta\mhalf (k) \label{fnlsig0} \\
\be
y(kL) \equiv \int^k_{L\mone} \frac {dp}{p} \( \frac p k \)^{\ts}
=
 \frac { 1- (kL)^{-\ts}}{\ts} \label{ydef}
.
\ee
The following limits apply
\be
y(kL) =
\left\{ \begin{array}{ll} \frac1\ts \qquad & (\ts \gg 1/\ln(kL))  \\
                                    \ln(kL) & (|\ts| \ll 1/\ln(kL)) \\
                \frac{(kL)^{|\ts|} }{|\ts|}  & (\ts \ll -1/\ln(kL)
                  \end{array} \right.
\label{yexp}
.
\ee

The  expressions for the 
correlators in terms of $y$ are;   
\bea
\calp_{\zetas} &=&  b^2 \calps +  \calp_{\delta\sigmas} \label{pzeta}  \\
\calp_{\delta\sigmas} &\simeq& 4y(kL)   \calps^2   \label{psofy}  \\
\calp_{\zetas} &\simeq& \calps \( b^2 + 4y \calps  \) \label{36a} \\
\frac35 \fnl &\simeq & 
 \frac{\calps^2}{\calpz^2} \( b^2 + 4y \calps  \)  \label{fnlofy1} \\
&\simeq & b^2 \frac{\calps^2}{\calpz^2} 
+  \frac12\sqrt{ \frac1{y(kL)} } 
 \( \frac{\calp_{\delta\sigmas}}{\calp_\zeta} \)^\frac32 
\frac1{\calp_\zeta\half} \label{fnlofy} \\
\tnl &\simeq& 
4   \frac{\calps^3}{\calpz^3} \( b^2 + 4y \calps  \)  \label{tnlofy1} \\
&\simeq &
4 b^2  \frac{\calps^3}{\calpz^3} +
   \frac1{y(kL)}  \( \frac{\calp_{\delta\sigmas}}{\calp_\zeta} \)^2 
\frac1{\calp_\zeta} \label{tnlofy}
.
\eea

The tilt of $\zeta$ is given by
\be
t_\zeta =\frac{
t_{\zetag} \calp_{\zetag} + b^2 \ts\calps
 + t_{\delta\sigmas} \calp_{\delta\sigmas} } {\calp_\zeta} 
\label{tzeta}
,
\ee
with 
\be
t_{\delta\sigmas}= \left\{ \begin{array}{ll} 2\ts \qquad & (\ts \gg 1/\ln(kL)) \\
                                     1/\ln(kL) & (|\ts| \ll 1/\ln(kL)) \\
                                     \ts & (\ts \ll -1/\ln(kL)
                  \end{array} \right.
\label{tsigmas}  
\ee

For zero or negative tilt, increasing the box size has an ever-increasing 
effect on the correlators. For positive tilt we can use a maximal box 
such that  $\ln(k L\sub{max}) \gg 1/\ts$ and $y=1/\ts$ are good 
approximations, giving
\bea
\calp_{\delta\sigmas} &\simeq& \frac4\ts  \calp_{\sigma}^2  \label{psofts}  \\
\frac35 \fnl &=& b^2 \frac{\calps^2}{\calpz^2} 
+  \frac12 t_\sigma\half
 \( \frac{\calp_{\delta\sigmas}}{\calp_\zeta} \)^\frac32 \frac1{\calp_\zeta\half}
\label{fnlofts} \\
\tnl &=& 4 b^2  \frac{\calps^3}{\calpz^3} +
   \ts \( \frac{\calp_{\delta\sigmas}}{\calp_\zeta} \)^2 
\frac1{\calp_\zeta}
\label{tnlofts}
.
\eea

%\bea
%\calpss(k) &\simeq&  4 \calp^2_\delta\sigma(k) (kL)^{|\ts|}/|\ts|  \\
%\tss &\simeq&  3\ts \\
%\frac35f\sub{NL}^\delta\sigmas &\simeq& 
%\frac {|\ts|\half}{\sqrt 2 (kL)^{\ts/2}} 
%\( \frac{\calpss(k)}{\calp_\zeta(k)}
%\)^\frac32 \calp_\zeta\mhalf (k) \label{fnlsig0} 
%.
%\eea

%\bea
%\calpss &\simeq & \frac4{\ts} \calp_{\delta\sigma}^2(k) \\
%\tss &\simeq& 2 \ts \\
%\frac35f\sub{NL}^\delta\sigmas &\simeq&  \frac{ \ts\half }{\sqrt 2}
%\( \frac{\calpss(k)}{\calp_\zeta(k)} \)^\frac32 
%\calp_\zeta\mhalf(k) \label{fnlsig}
%.
%\eea

\subsection{Working in a minimal  box}
\label{three}

To minimize the cosmic uncertainty of the correlators, one might wish to choose
 the box size  to be 
 as small as possible, consistent with the 
condition $LH_0\gg 1$ which is required so that it can describe the whole
observable Universe \cite{myaxion}. 
How big $LH_0$ has to be depends on the accuracy required
for the calculation of observables, using the curvature perturbation as the 
initial condition. As the  equations required for that calculation
 involve $k^2$ rather than 
$k$ it may be reasonable to suppose that very roughly
$1\%$ accuracy will be obtained with $LH_0\sim 10$ 
and $0.01\%$ accuracy with $LH_0\sim 100$. 
Even with the latter, the minimal box size $L\sub{min}$ corresponds only
to 
\be
\ln (L\sub{min}  H_0) \simeq 5
\label{minimal}
.
\ee

The range  of cosmological scales is usually taken to be only
 $\Delta \ln k\simeq 14$,
going from the size $H_0\mone\sim 10^4\Mpc$ of the observable Universe,
to the scale $10\mtwo\Mpc$ which encloses  a mass of order $10^6\msun$
and which corresponds to the first baryonic objects.

Cosmological scales therefore correspond to roughly
\be
\ln (kL\sub{min})\sim 5\ {\rm to}\  20
.
\ee

Let us see how things work out with the  minimum box size.
We saw that the dependence on the box size is through the function
$y(kL)$ given by \eqs{ydef}{yexp}. With the minimal box, this
function is of order 1  on all cosmological scales,
provided that  $|\ts| \lsim 1/20$.  
This bound on $\ts$
is more or less demanded by observation if $\zetag$ is negligible, but it
can be far exceeded if $\zetag$ dominates. 
In the latter case
I will allow only positive tilt, since strong negative tilt looks 
unlikely as seen in Section \ref{3g}. Then $y$ 
is at most $1/\ts$, and hence still roughly of order 1.

%The  observational bounds on  $|\fnl|$ and $\tnl$  imply 
%\bea
%|f\sub{NL}|\calp_\zeta\half  &\ll& 1 \label{fnlbound2} \\
%\tau\sub{NL}\calp_\zeta &\ll& 1 \label{tnlbound2}
%.
%\eea
%{}From the second terms of \eqs{fnlofy}{fnlofy} 
%one sees that for a minimal box size,
%corresponding to $y\sim 1$, each of 
%\eqsref{fnlbound2}{tnlbound2} implies
%\be
%\rng \equiv \( \frac{\calp_{\delta\sigmas}}{\calp_\zeta} \)\half \ll 1
%\label{rngdef}
%.
%\ee 
%This is in line with the fact that the data are consistent with $\zeta$
%being perfectly gaussian. It means that the contribution of ${\delta\sigma}^2$
%to the curvature perturbation is sub-dominant.

To go further with the minimal box, it 
will be enough to consider the two extreme cases, that the correlators are 
dominated by either their `linear'  or their `quad' contributions.
{}From \eqss{36a}{fnlofy1}{tnlofy1} the former case corresponds to
\be
\calps \ll b^2
\label{sigcon}
.
\ee
Given \eq{specinterp}, this corresponds to the linear term of $\zetas$ 
dominating, while the opposite case corresponds to the quadratic term of
$\zetas$ dominating.

If the linear term dominates and 
 $\zetag$ is negligible,
\bea
\calpz &=& b^2\calps \\
\frac35 \fnl &= & b\mtwo 
%= \rng \calpz\mhalf 
\label{55} \\
\tnl  &=& (36/25) f\sub{NL}^2
.
\eea
This is  the usually-considered case. With a change of 
 normalization  one can write   \cite{spergel}
$\zeta=\delta\sigma+ \frac35\fnl {\delta\sigma}^2$.

If instead the quadratic term dominates $\zetas$, it cannot dominate $\zeta$
or there would be too much non-gaussianity.
 Indeed, 
given the interpretation \eqreff{specinterp}, the non-gaussian fraction
is
\be
\rng \equiv \( \frac{\calp_{\delta\sigmas}}{\calp_\zeta} \)\half 
=\( \frac{\calp_{\zetas}}{\calpz} \)\half
\label{rngdef}
, 
\ee 
and  \cite{bl}
\be
\rng \sim  \(  |\fnl| \calp_\zeta\half \)^\frac13 
\simeq   \(  \tnl \calp_\zeta \)^\frac14 
\label{rngexp}
.
\ee
The present bound $|\fnl|< 121 $ requires  $\rng <0.2$, but the present
 bound on 
$\tnl <10^4$ requires $\rng< 0.07$.
 We see that {\em if the quadratic
term dominates, the present bound on the trispectrum is a stronger constraint 
than the one on the bispectrum.} 
Absent a detection, the post-COBE bounds  on the bispectrum and trispectrum
will lead to about the same constraint,  $\rng\lsim 0.04$.

With the quadratic term dominating, the 
present bound on $\tnl$ gives  $\calp_{\zetas}\half
<3\times 10^{-6}$. This, though, is 
 on the fairly large  cosmological scales probed by the cmb anisotropy.
With positive tilt $\calp_{\zetas}(\kmax)$ could be much bigger,
even not far below 1 leading to black hole formation.

\subsection{Running the box size}
\label{two}

We have found that the stochastic properties 
 depend on the size
of the box in which \eqs{withlin}{vevs} are  supposed to hold.
%This factor arises from the infrared divergence of \eq{phisk},
%which in turn is due to the interference of very long wavelength components.
%The logarithmic divergence corresponds to our assumption that $\calps$
%is perfectly scale-independent; it will be more severe if $\calps(k)$
%decreases with $k\mone$ and absent (in principle) in the opposite case.
This seems to be incompatible with a basic tenet of physics concerning the use
of Fourier series, that the box size should 
be irrelevant if it is much bigger than the scale of interest.

This situation was discussed in \cite{bl} on the assumption that $\calps$
is scale-independent, and it is easy to extend the discussion to the case
of an arbitrary $\calps(k)$. 
 The crucial point is that  $\overline{\delta\sigma}$ is supposed to
vanish {\em within the chosen  box} of size $L$. 
Let us imagine 
now that this box is within a much bigger box of fixed
size $M$, and see how things vary if the size and location of the smaller
box are allowed to vary. We have in mind that the small box will be a 
minimal one, and in the case of constant positive tilt the big
 box might be the maximal one satisfying $\ln(MH_0)\ts \gg 1$.
Defined in the big box, $\zeta$ has the form \eqreff{withlin},
with some coefficient $b$ and with   $\overline{\delta\sigma}=0$.

Focus first on a particular box with size $L$,
and denote quantities evaluated  inside this box
 by a subscript $L$. In general  $\overline{\delta\sigma}_L$ 
will not  vanish, and absorbing its  expectation value  into $b$ 
using \eq{beq} we find   
\be
b_L = b + 2{\overline{\delta\sigma}}_L
.
\ee
Now,  instead of considering a particular  small box, let its
 location vary so that
 $\overline{\delta\sigma}_L$ becomes the original perturbation $\delta\sigma$
smoothed on the scale $L$. Then
\bea
\vev{b^2_L} &=& b^2 + 4 \vev{ \overline{\delta\sigma^2}_L } \label{vevb}
\nonumber \\
\vev{\overline{\delta\sigma^2}_L} &=&  \int_{M\mone}^{L\mone} 
\frac {dk}k \calps(k) 
\label{lvev} 
,
\eea
where the expectation values refer to the big box.

The operations of smoothing and taking  the expectation value
commute. Therefore,
if $\calp^L_{\delta\sigmas}$,  $B_{\delta\sigmas}^L$ and $T_{\delta\sigmas}^L$ 
are the spectrum,  bispectrum and trispectrum
defined within a particular small box of
size $L$, we should have
\bea
\vev{\calp^L_{\delta\sigmas}} &=& \calp_{\delta\sigmas} \label{1} \\
	     \vev{B_{\delta\sigmas}^L} &=& B_{\delta\sigmas} \label{2} \\
\vev{T_{\delta\sigmas}^L} &=& T_{\delta\sigmas} \label{3}
,
\eea
where the right hand sides and the expectation values refer to the big box.
One can verify this explicitly 
 using \eqss{specfirst}{bfirst}{tfirst}. Indeed, these equations
apply to a box of any size.
The hierarchy $k\gg L\mone \gg M\mone$ 
allows one to evaluate the changes in the 
integrals  induced by the  change $M\to L$, 
and to verify (for any form of the spectrum) 
that this change is  precisely compensated by  the change
$b^2\to \vev{b^2_L}$. 

\section{The inflationary prediction}
\label{four}

In this section we see how \eq{withlin} 
and its generalizations may be predicted by
 inflation. We take the relevant scalar fields to be canonically normalized,
and focus mostly on slow-roll inflation with Einstein gravity. The latter
restriction is not very severe because a wide class of non-Einstein
gravity theories can be transformed to an `Einstein frame' \cite{book}.

\subsection{The curvature perturbation}

The curvature perturbation $\zeta$ is defined on the spacetime slicing 
of uniform energy density $\rho$ through the metric 
 \cite{sb,lms,maldacena,cz,rst}
\be
g_{ij}= a^2(t) e^{2\zeta(\bfx,t)} \gamma_{ij}(\bfx)
\label{zetadef}
,
\ee
where $\gamma$ has unit determinant. 
Within the inflationary cosmology,  if the tensor perturbation is 
negligible, $\gamma_{ij} = \delta_{ij}$.

By virtue of the separate universe
assumption, the  threads of  spacetime orthogonal to the uniform density
slicing are practically comoving. As a result, 
a comoving volume element $\calv$ is proportional
to $a^3(\bfx)$ where
\be
 a(\bfx,t) \equiv  a(t) e^{\zeta(\bfx,t)}
\label{zetadef2}
.
\ee
This  means that $a(\bfx,t)$ is 
 a locally-defined scale factor.

An alternative definition of $\zeta$ refers to 
 the slicing where the metric has the form  $g_{ij}
=a^2 \hat\gamma_{ij}$ (with again $\hat \gamma$ having unit determinant).
This is usually  called the spatially flat slicing,
which it is if the tensor perturbation 
is negligible. Linear cosmological perturbation theory gives
$\zeta$ in terms of  $\delta\rho$ on the spatially flat slicing,
\be
\zeta = -H\frac{\delta\rho}{\dot\rho} = \frac{\delta\rho}{3(\rho + P)}
\label{zetaofdrho}
.
\ee 
If $\fnl$ turns out to be of order 1 though, it will be necessary to go
to second order and then the non-perturbative 
definition \eqreff{zetadef} becomes  more useful  \cite{lr,lrfirst}.

The linear expression \eqreff{zetaofdrho} is convenient if the fluid
is the sum of fluids, each with its own $P(\rho)$. Defining on flat slices
the constants
\be
\hat\zeta_i \equiv \frac{\delta\rho_i}{3(\rho_i + P_i)}
,
\ee 
one has
\be
\zeta(t) = \frac{\sum (\rho_i + P_i) \hat\zeta_i}{(\rho  + P)}
\label{zetamulti}
.
\ee

By virtue of the separate universe
assumption,  the change in the energy $\rho \calv$ within a given 
comoving volume element is equal to 
 $-Pd\calv$ with $P$ is the pressure. This
 is equivalent to the continuity equation
\bea
\dot\rho &=& - 3 \frac{ \dot a(\bfx,t) }{a(\bfx,t)}
\( \rho + P \) \\
\frac{\dot a(\bfx,t)}{a(\bfx,t)} &=& \frac{\dot a}a
+\dot\zeta
.
\eea
Remembering that this holds on
 uniform-density slices, we see  that
 $\zeta$ is  conserved 
 \cite{bardeen,bst,my85,lms,cz,rst,lvcons} during any era when $P$
is a unique function of $\rho$. That is guaranteed
 during any era where there is
 practically 
complete radiation domination ($P=\rho/3$),  matter domination
($P=0$) or kination ($P=\rho$). The generation of 
 $\zeta$ may take place during any other era.

\subsection{Generating the curvature perturbation}

The idea is that during inflation,
the vacuum fluctuation of each light field becomes, a few Hubble times
{\em after}  horizon exit, a classical perturbation. (To keep the language 
simple I shall loosely say that the classical perturbation is present
{\em at} horizon exit.) 
 The observed curvature perturbation, present a few Hubble times {\em before}
cosmological scales start to enter the horizon, 
is generated by one or more of these classical field perturbations.

 Opportunities for generating $\zeta$ occur during any era when there is no 
relation $P(\rho)$. Generation 
 was originally  assumed to take place promptly at horizon
exit in a single-component inflation model \cite{starob82,singlecom,bst}. 
Then it was realised  \cite{starob85} 
that in a multi-component inflation model there will be continuous
 generation during inflation, but  it was still assumed that the curvature
perturbation achieves its final value by the end of inflation.

The term `curvaton-type models' in this paper denotes models in which
a  significant contribution to the curvature perturbation is generated after
the end of slow-roll inflation, by the perturbation in a field which has
a negligible effect on inflation. If the curvaton-like 
contribution is completely dominant then the 
mechanism of inflation is irrelevant. 

The curvaton model itself was the original proposal. 
In this model,  the
 oscillating curvaton field leads to a second reheating, and 
the curvature perturbation is caused by the perturbation in the
curvaton field
 \cite{mollerach,lm,curvaton,moroi,luw}.\footnote
{Earlier papers \cite{precurvaton} considered the  generic scenario, in which
  a light scalar field  
gives a negligible contribution 
to the energy density and  the curvature perturbation during inflation, 
but a significant one at an unspecified  later epoch. Such a scenario becomes
a curvaton-type model if that epoch is before cosmological scales start
to leave the horizon,  otherwise it may be  an axion-type 
 model giving a cdm isocurvature perturbation.
Among curvaton-type models, the  curvaton model is  the one which  generates
the curvature perturbation from the
 perturbation in the  amplitude   of the oscillating
curvaton field. It was described in
 \cite{mollerach}, and a formula equivalent
to the estimate $\zeta\sim \delta\sigma/\sigma$ was given in \cite{lm}.
In \cite{curvaton} the curvaton model was advocated as the dominant cause of
the curvature perturbation and a 
 precise calculation of $\calpz$ was made 
 allowing for significant radiation. In 
\cite{moroi}  a significant  inflaton contribution was allowed.
The first calculation
of $\fnl$ was given in \cite{luw}. The curvaton mechanism with a 
pre-big-bang instead of inflation was worked out in \cite{es}.}
Alternatives to the curvaton model, which still use a reheating, are
to have the curvature perturbation generated by an inhomogeneity in 
any or all of the decay rate \cite{hk,varcoupling},
  the mass   \cite{varmass} or 
the interaction rate \cite{thermalization}
of the particles responsible for the reheating. In that case the 
reheating can be the first one  (caused by the scalar field(s) responsible
for the energy density during inflation) or alternatively the particle
species causing the reheating can be a fermion  \cite{bc}.
Other opportunities for generating the curvature perturbation occur
 at the end of inflation \cite{endinflation}, during
preheating \cite{preheating}, and 
 at a phase transition producing cosmic strings \cite{matsuda}.

To describe the generation of the curvature perturbation
 I make two assumptions. First, that
the evolution of perturbations on a given cosmological scale $k$
can be described using an idealized universe, which is smooth on some scale
a bit smaller than $1/k$. Second, that the local evolution of this idealized
universe in the super-horizon regime 
can be taken to be that of some unperturbed universe.

The first assumption is routinely made in cosmology, on both super- and
sub-horizon scales. The second assumption
is the separate universe assumption \cite{bst,sb,llmw},
 which amounts to the statement that
the smoothed universe becomes  locally isotropic and homogeneous when the 
smoothing scale is much bigger than the horizon. 
Given the content of the 
Universe at a particular epoch, it may be checked using  cosmological
perturbation theory,  or else using the gradient expansion 
\cite{sb,Deruelle,shsa,lms}
with the additional
assumption of local isotropy. Local isotropy is more or less \cite{bh} 
guaranteed \cite{wald} 
by inflation. Independently of particular considerations, the separate
universe assumption has to be valid on a scale a bit  larger than
 $H_0\mone$, or the concept  of an unperturbed FLRW Universe
would make no sense. It will  therefore be valid on all cosmological
scales provided that all relevant scales in the early Universe are much
smaller, which will usually be the case \cite{lwcons}.

I assume  slow-roll inflation, noting possible generalizations along the way.
The  light fields  $\phi_i$ are   defined as those  which satisfy
 flatness conditions;
\be
\epsilon_i \ll 1 \qquad |\eta_{ij}|\ll 1
\label{59}
\,,
\ee
where
\bea
\epsilon_i &\equiv & \frac12 \mpl^2 \( \frac{V_i}{V} \)^2 \label{60} \\
\eta_{ij} &\equiv& \mpl^2 \frac{V_{ij}}{V} \label{61}
,
\eea
with $V_i\equiv \pa V/\pa \phi_i$ and $V_{ij}\equiv \pa^2 V/\pa\phi_i 
\pa\phi_j$.

The exact field equation for each light field,
\be
\ddot \phi_i + 3H\dot\phi_i + V_i =0
\label{62}
,
\ee
  is supposed to be well-approximated by 
\be
3H\dot\phi_i=- V_i
\label{63}
.
\ee

In these expressions $H$ is the Hubble parameter, related to 
$V$ by
\be
3\mpl^2 H^2 = V + \frac12 \sum \dot\phi_i^2
\label{64}
.
\ee 
By virtue of \eqss{59}{60}{63},
 this becomes $3\mpl^2 H^2 \simeq V$,
and $H$ is slowly-varying corresponding to almost-exponential
inflation;
\be
\left| \frac1{H^2}\frac{ d H}{d t}  \right| \simeq 2\epsilon
\ll 1
,
\ee
with
\be
\epsilon \equiv \sum \epsilon_i
.
\ee
(Despite the notation, it is $\sqrt \epsilon_i$ and not $\epsilon_i$
which transforms as a vector in field space.)
If inflation   is almost exponential but not necessarily  slow-roll,
it may  still be useful to define the light fields by
\eqst{59}{61}, with $V$ in the denominator replaced by
$3\mpl^2H^2$.

Focusing on a 
 given epoch during inflation, it may be
 convenient to choose the  field basis
so  that   one field
$\phi$ points along the inflationary trajectory. I will call it the inflaton,
which coincides with the standard terminology in the case of 
single-component inflation.
I will denote the 
orthogonal light fields (assuming that they exist) by $\sigma_i$.
In an obvious notation $\epsilon_{\sigma_i}=0$ and
$\epsilon = \epsilon_\phi$ initially.  One may define also
$\eta\equiv \eta_{\phi\phi}$. 

{}From  \eqst{59}{61} and \eqreff{63},
 the gradient of the potential is slowly
varying;
\be
\frac1H\frac{d V_i}{dt} =  - \eta_{ij} V_j
.
\ee
The possible inflationary
trajectories are the lines of steepest descent of the potential.
The trajectories  may be practically straight (single-component inflation) or
 significantly curved in the subspace of two or more light fields
(multi-component inflation, called double inflation in the case of 
two fields).  The terms single- and multi-component
 refer to the viewpoint that
the inflaton field is a vector in field space. For single-component
inflation, $\epsilon_{\sigma_i} \ll \epsilon$, and the
 inflaton field
$\phi$ hardly  changes direction in field space
, so that one can choose a
practically fixed 
 basis $\{\phi,\sigma_i\}$.

\subsection{The $\delta N$ formula}

To evaluate the curvature perturbation generated by the vacuum fluctuations
of the light fields,  we can use the
$\delta N$ formalism  \cite{starob85,ss,lms,lr,st}.
It gives $\zeta(\bfx,t)$, in terms of the light field perturbations defined 
on a flat slice at some  fixed `initial' epoch during inflation;
\be
\phi_i(\bfx) = \phi_i + \delta\phi_i(\bfx)
\label{phipert}
.
\ee
 Keeping terms which are linear and quadratic in $\delta\phi_i$,
 the
time-dependent curvature perturbation is
\bea
\zeta(\bfx,t)  &=& \delta N(\phi_i(\bfx),\rho(t)) \nonumber\\
&=& \sum_i N_i \delta\phi_i(\bfx) + \frac12 \sum_{ij}
N_{ij} \delta\phi_i \delta\phi_j
\label{zetaofdphi}
\eea
Here, $N(\phi_i,\rho)$ is the number of $e$-folds, evaluated in an
unperturbed universe, from the initial epoch 
 to an  epoch when the energy density $\rho$ has a specified
value. 
In the  second line, $N_i\equiv \partial N/\partial \phi_i$
and  $N_{ij}\equiv \partial^2 N/\partial \phi_i\pa \phi_j$, both evaluated on
the unperturbed trajectory. In  known
cases  the first two terms of this expansion in the field perturbations
are  enough.

%In the following  I  take the $\delta\phi_{i*}$ to be
%    gaussian and uncorrelated. Then \eq{zetaofdphi} allows the correlators
%of $\zeta$ to be calculated. Indeed, 
%choosing the 
%field basis and the field normalization so that $N_{ij}=
%\delta_{ij}$, the curvature  perturbation is of the form
%\be
%\zeta = \sum \( b_i \delta\sigma_i + (\delta\sigma_i)^2 \)
%,
%\ee
%and each  correlator is the sum of contributions given by
%\eqst{specfirst}{tsigs}.

The curvature perturbation $\zeta(\bfx,t)$ is independent of the 
`initial' epoch. At that epoch let us work in a basis $(\phi,\sigma_i)$
so that $N_i=(N_\phi,N_{\sigma_i})$. 
%This epoch determines the 
% flat spacetime slice on which are defined
% the perturbations $\delta\phi_i(\bfx)$, and it provides the initial 
%epoch from which the number $N(t)$ of $e$-folds is calculated.
The contribution of the inflaton to $\zeta$ is
 time-independent because 
 $\delta\phi$ 
  just corresponds to a shift along the inflaton trajectory.
It  is given by
\bea
\zetag &=&   N_\phi \delta\phi + \frac12 N_{\phi\phi} (\delta\phi)^2 \\
N_\phi &=& \frac1{\mpl^2} \frac{V}{V'} \label{nphi} \\
N_{\phi\phi}/N_\phi^2  &=& \eta - 2\epsilon \label{nphiphi}
.
\eea
Redefining $\phi$  we learn
that the non-gaussianity coming from $\delta\phi^2$ 
is negligible ($|\fnl|\ll 1$).\footnote
{The  three-point
correlator of  $\delta\phi$ also \cite{sl,zrl}  gives $|\fnl|\ll 1$.
 A different proof that single-component inflation gives $|\fnl|\ll 1$
 was given earlier by Maldacena \cite{maldacena}.}
We can therefore write
\be
\zetag \simeq    N_\phi \delta\phi
\label{zetasingle}
.
\ee

The   fields $\sigma_i$
correspond to shifts orthogonal to the
inflaton trajectory. The perturbations $\delta \sigma_i$
 give initially 
 no contribution to the curvature perturbation; in other words the
$\sigma_i$ are {\em
initially} isocurvature perturbations. 
If the $N_{\sigma_i}$ were evaluated at a final epoch 
just a  few Hubble times after the initial one, they  would be practically
zero because the possible trajectories can be taken to be
practically straight and 
parallel over such a short time. The crucial point is that
the $N_{\sigma_i}$ can subsequently grow.
Such growth  has nothing to do with the situation at
horizon exit, in particular there is no expression for the
$N_{\sigma_i}$ in terms
of the potential gradient analogous to \eqs{nphi}{nphiphi}.

In any case,  $\zeta$ settles down at some stage to a final
time-independent value, which persists until cosmological scales start
to enter the horizon and is constrained by observation.
This value is the  one that we studied in the previous section,
focusing on \eq{withlin} which is obviously a special case
of \eq{zetaofdphi}. 

Although we focused on slow-roll inflation, the basic formula
\eqreff{zetaofdphi} can describe the generation of $\zeta$ from the 
vacuum fluctuation for any   model of almost-exponential inflation, which
 need have nothing to do with scalar fields or Einstein gravity.
The only requirement is that the specified light fields $\phi_i$,
satisfying \eqst{59}{61} 
with $V$ in the denominator replaced by $3\mpl^2 H^2$, determine the 
local evolution of the energy density and pressure 
 until the approach of
horizon entry. Going further, the $\phi_i$ need not 
be the light fields themselves,
but  functions of them, evaluated at an `initial' epoch which might be after
the end of inflation.

\subsection{The gaussian approximation}

In the limit where the potential of the light fields is perfectly flat,
inflation is exponential (de Sitter spacetime) with constant $H$.
At least up to second order in cosmological perturbation theory
\cite{sl}, 
the perturbations in the light fields are then generated from the vacuum 
without back-reaction.  Each  perturbation is classical, and a few Hubble
 after horizon exit is  practically time-independent with the 
spectrum \cite{bd} $(H/2\pi)^2$. 
 Keeping a slow variation of $H$ and
finite values of the parameters $\epsilon_i$
and $\eta_{ij}$, the 
 three-point correlator of the light field perturbations
has  been calculated by Seery and Lidsey
\cite{sl}, and its  effect has been shown indeed to be negligible
 in some particular cases \cite{sl,lz,zrl}. In the following we take the
light fields to be gaussian at horizon exit.

Since $\zeta$ is almost gaussian it will presumably be a good approximation to
keep only the first  term
 of \eq{zetaofdphi}. Let us recall briefly the situation for that case.
Focusing on a particular scale, we can take the initial epoch to be the epoch
of horizon crossing. (The horizon-crossing trick.) Then
\be
\zeta(\bfk) = \sum_i N_{i*} \delta\phi_{i*}(\bfk)
\label{zetahorcross}
,
\ee
where a star denotes the epoch of horizon exit.
Using the summation convention,
\be
\calp_\zeta(k)  =  (H_*/2\pi)^2 N_{i*}N_{i*} 
\label{multispec}
.
\ee

To evaluate the tilt from this equation
one may use the slow-roll expressions
\bea
\frac d{d\ln k} &=& -\frac{\mpl^2}{V} V_i \frac{\pa}{\pa \phi_i} \\
N_i V_i &=& \mpl\mtwo V
,
\eea
to give \cite{ss,treview}
\be
t_\zeta
 =  \(
2\frac{\eta_{nm} N_nN_m}{N_iN_i} -2\epsilon -\frac2{\mpl^2 N_iN_i} 
\)_*
\label{multitilt}
,
\ee
with  the right hand side evaluated at horizon exit.

At horizon exit, let us use the basis $\phi,\sigma_i$. 
 If the inflaton  contribution $\delta\phi$ dominates 
one recovers the standard  predictions;
\bea
\calp_\zeta &=& \frac 1{2\mpl^2\epsilon_*} \( \frac{H_*}{2\pi} \)^2 
\label{pinf} \\
 t_\zeta &\simeq & t_\phi= \( 2\eta- 6 \epsilon \)_*
\label{tphi}
.
\eea
In this case, the tensor fraction $\calp\sub{t}/\calpz$ is $r=16\epsilon$.

The contributions $\delta\sigma_i$ to $\calpz$ are positive making
\cite{treview} $r\leq 16\epsilon$. If one or more of them completely 
dominates, $r$ will  be too small to ever observe. 
 If just one of them dominates (call it $\sigma$) dominates,   
\bea
\calp_\zeta &=& N_{\sigma *}^2  \( \frac{H_*}{2\pi} \)^2  \\
t_\zeta &\simeq &  t_\sigma= \( 2\eta_{\sigma\sigma} - 2\epsilon \)_*
\label{tsigma}
.
\eea

\subsection{Non-gaussianity in curvaton-type models}

Now I consider  the quadratic term of \eq{zetaofdphi}, 
allowing for the first
time the possibility of significant scale dependence.

%Since observation is consistent with $\zeta$ being gaussian, the 
%quadratic term presumably gives a negligible contribution to the 
%spectrum so that the latter is still given by \eq{multispec}.
%(We demonstrated this
%for the special case considered in the previous  section.)
%The importance of the quadratic term is that it can  generate 
%\cite{lr} observable non-gaussianity.

The horizon-crossing trick for evaluating scale
dependence may  not work when the quadratic term is included, 
because  the convolution \eqreff{phisk} involves a range
of wavenumbers.  It can still be used if the correlators of interest involve
only a narrow range of scales $k_i\sim k$, and  one just wishes to 
evaluate the dependence on the overall scale $k$.
Otherwise, one should take 
 the `initial'  epoch for  \eq{zetaofdphi}  to  be a fixed one, 
after all relevant scales have left the horizon. 

To handle this situation I suppose that 
inflation
is single-component, and that there is
 just one  non-inflaton field $\sigma$ which is going to be the curvaton-like
field. 
 (With small modification the discussion still applies
to  multi-component inflation models generating negligible
non-gaussianity.) Then
\bea
\zeta &=& \zetag + \zetas \\
\zetag &\simeq & N_\phi \delta\phi \label{94} \\
\zetas &=& N_\sigma \delta\sigma 
+ \frac12N_{\sigma\sigma} {\delta\sigma}^2
\label{95}
.
\eea
Up to  the normalization of $\delta\sigma$, \eq{95} is the same as
\eq{withlin}.

Since $\epsilon_\sigma\ll \epsilon$, the perturbations on flat slices satisfy
 \cite{tn}
\bea
\frac1{H}\frac{d\delta\phi}{dt} &=& -\( \eta-2\epsilon \) 
 \delta\phi  \\
\frac1{H}\frac{d\delta\sigma}{dt} &=& -\eta_{\sigma\sigma} \delta\sigma 
\label{sigev}
,
\eea
The second equation is just the first-order perturbation of the
 unperturbed slow-roll equation $3H\dot\sigma = - V_\sigma$. That equation
therefore applies locally.

The solutions are
\bea
\delta\phi = e^{-\int^t_{t_*} H (\eta-2\epsilon) dt} \delta\phi_* \\
\delta\sigma = e^{-\int^t_{t_*} H \eta_{\sigma\sigma} dt} \delta\sigma_*
.
\eea
These are to be evaluated at a fixed $t$ which
will be the `initial' epoch for use in \eqs{94}{95}.
Keeping only the linear terms and evaluating the spectrum, we recover 
\eqs{tphi}{tsigma} for the spectral indices.

{}From now on the focus will be on the quadratic 
 potential
\be
V(\sigma) = \frac12m_*^2\sigma^2 
\label{70}
.
\ee
Also  $H$ is supposed to  be sufficiently slowly varying
that
\be
\ts\simeq 2\eta_{\sigma\sigma} \equiv \frac{2m_*^2}{3H_*^3}
,
\ee
where $H_*$ is now the practically constant value of $H$ during inflation
without reference to a particular epoch.

As with  any situation involving
scalar fields in the early Universe,  one has to remember
that  the
 effective potential
$V(\sigma)$ can  be affected by the values of other scalar fields and change
with time. In particular,  supergravity gives for a {\em generic}
field during inflation $|\eta_{\sigma\sigma}|\sim 1$. This 
marginally violates the slow-roll condition and corresponds
roughly to $m_*\sim H_*$ and $\ts\sim 1$.
 The  field which dominates
$\zeta$ must have   $|\eta_{\sigma\sigma}|\lsim 0.01$.
In any case, the mass $m_*$ appearing in \eq{70} will generally not be the 
true mass $m$, defined  in the vacuum.

\subsection{Maximum wavenumber and the  `initial' epoch}

The classical curvature perturbation $\zeta$ is generated  up to some maximum
wavenumber $\kmax$. This maximum is generally taken to correspond to a scale
$1/\kmax$ far below the shortest scale of direct cosmological
interest discussed in Section \ref{three}. Nevertheless the value of
$\kmax$ matters. It represents the shortest possible
scale for the  formation of primordial 
 black holes\footnote
{See for instance references
in \cite{lmsz}, where the quantum regime $k>\kmax$  is also considered.}
and the shortest scale on which matter density perturbations can exist.
As we shall see, short-scale perturbations in the  curvaton density
contribute to its mean density, and hence indirectly to the magnitude of
$\zeta$ which the curvaton model generates.

If $\zeta$  is created during inflation, $\kmax$ is the scale $k\sub e$
leaving the horizon at the end of inflation. It is given by
\be
\frac{k\sub e}{H_0} = e^N
\label{keofN}
,
\ee
where $N$ is the number of $e$-folds of slow-roll inflation after the 
observable Universe with present size $H_0\mone$ leaves the horizon.
For a high inflation scale with continuous radiation domination afterward
 $N\simeq 60$. To make  $k\sub e\mone$ 
comparable with  the shortest cosmological
 would require $N\simeq 14$ (see the discussion after \eq{minimal}) which is 
hard to achieve. For this reason it seems to have been assumed 
in all previous discussions that $k\sub{max}\mone$ 
will be far below cosmological
scales.

That assumption is not justified if $\zeta$ is created by a curvaton-type
mechanism long after inflation. 
After inflation, the perturbation $\delta\sigma$ in the curvaton-type field
redshifts away on scales entering the horizon. Therefore, $\kmax$ is the 
scale entering the horizon when $\zeta$ is created. 
(See \cite{myaxion} for the same phenomenon in the axion case) This scale
leaves the horizon long before the end of inflation. To handle that situation,
I will still equate  $\kmax$ with $k\sub e$ given by \eq{keofN}, but define
$N$ as the number of $e$-folds of {\em relevant} inflation after the observable
Universe leaves the horizon, `relevant' meaning $e$-folds which produce
perturbations on scales bigger than $\kmax\mone$. Demanding only that 
$k\sub{max}\mone$
is below the shortest cosmological scale we can allow a range
\be
14 < N \lsim 60
.
\ee
Taking the extreme values $N=60$ and (say) $\ts=0.4$ gives
$e^{N\ts}\sim 10^{11}$.

The end of relevant inflation is the appropriate `initial' epoch for 
use in \eq{95}.
The spectrum of $\sigma\sub e$ is
\be
\calp_{\sigma\sub e} = \( \frac{H_*}{2\pi} \)^2 \( \frac{k}{k\sub e} \)^{\ts}
= \( \frac{H_*}{2\pi} \)^2 e^{-N\ts} \( \frac{k}{H_0}  \)^{\ts}
\label{psige}
,
\ee
The  factor $e^{N\ts}$
is of order 1 if $N\ts \lsim 1$.
 This is more or less demanded by the observational bound on $t_\zeta$
if the curvature perturbation is dominated by the curvaton contribution.
In the opposite case large tilt is  allowed, making 
$e^{N\ts}$ exponentially large as we saw earlier.

The unperturbed field at the end of inflation is 
\bea
\sigmae^2 &=& \sigma^2 e^{-N_L \ts} \label{sigeofsigl} \\
N_L &\equiv & \ln(k\sub e L) \label{nldef} 
,
\eea
where $\sigma$ is the practically unperturbed value of the curvaton field
within the box of size $L$ when it leaves the horizon, and $N_L$ is the number
of $e$-folds of relevant inflation after the box leaves the horizon.
 
\subsection{Cosmic uncertainty}
\label{3g}

If inflation lasts for enough  $e$-folds before the observable Universe
leaves the horizon, the stochastic formalism \cite{stochastic} allows
one to calculate the probability distribution of $\sigma $, for a random
location of our Universe. If our location is typical the actual value
of $\sigma ^2$ will be roughly $\vev{\sigma ^2}$.

If the variation of $H$ is  so slow that it can be ignored, one arrives
at a particularly simple probability distribution which is described 
in this subsection.
Generalizations to allow for varying $H$ 
 are given for instance in \cite{ouraxion,lm2}.

\subsubsection{The Bunch-Davies case}

In the case of constant positive tilt the result can be obtained 
from the formalism already presented. To do this, one works in a maximal box
and assumes  that 
the average  value of $\sigma$ rolls down to a practically zero value
well before the observable Universe leaves the horizon. At any subsequent
epoch, the local value  $\sigma(\bfx)$ has a gaussian probability distribution
with variance
\bea
\vev{\sigma ^2(\bfx)} &=& \( \frac{H_*}{2\pi} \)^2 \int^{aH_*}_0 \frac {dk}k
\( \frac k{aH_*} \)^{\ts} \label{bdvev1}\\
& = &  \( \frac{H_*}{2\pi} \)^2 \frac1{\ts} =\frac{3H_*^4}{8\pi^2 m_*^2}
\label{bdvev}
.
\eea
This is the case considered by Bunch and Davies \cite{bd,starob82}.

Working within a smaller box with size $L$ (thought of as a minimal one), 
$\sigma(\bfx)$ has a spatial average and a perturbation;
\be
\sigma(\bfx) = \sigma_L + \delta\sigma_L(\bfx)
\label{sat1}
,
\ee 
where the classical perturbation includes all wavenumbers $k<aH$.
The  mean square within that box is
\bea
\vev{\sigma^2(\bfx)}_L &=& \sigma_L^2 + \vev{\delta\sigma_L^2}_L \\
&=& \sigma_L^2 + \int^{aH_*}_{L\mone} \frac{dk}{k} \calp_\sigma
.
\eea
For a random location of the small box (within the maximal box)
 each term of \eq{sat1}
has a gaussian distribution.
Adding the two variances gives
\be
\vev{\sigma^2(\bfx)} = \int^{L\mone}_0 \frac{dk}{k} \calp_\sigma
 + \int^{aH_*}_{L\mone} \frac{dk}{k} \calp_\sigma
\label{fr1}
,
\ee
which agrees with \eq{bdvev1}.

{\em When the minimal box first leaves the horizon} the perturbation
$\delta\sigma$
is negligible.  For a random location of the minimal box, 
the  variance of  the unperturbed value $\sigma$
is then  practically equal to the Bunch-Davies expression
\eqreff{bdvev}. 

We were defining the curvature perturbation \eqreff{95}
within a  minimal box, because that  has general applicability.
In the Bunch-Davies case we can instead use  a maximal box, big enough to 
ensure the condition $\sigma=0$ before the observable Universe leaves the
horizon.
In that case, 
$N_\sigma$ may vanish 
leaving only the quadratic term of \eq{95}. This will happen
if $\sigma\to -\sigma$ is a symmetry of the theory, and it happens anyway
in the actual curvaton model because $N_\sigma$ is then determined directly
by the potential. Then  the correlators are given by 
\eqst{psofts}{tnlofts}, with $b=0$ and 
$\calp_{\delta\sigma^2}=\calp_{\zetas}$;
\bea
\calp_{\zetas} &\simeq& \frac4\ts  \calp_{\sigma}^2  \label{psofts2}  \\
\frac35 \fnl &=&   \frac12 t_\sigma\half
 \( \frac{\calp_{\zetas}}{\calp_\zeta} \)^\frac32 \frac1{\calp_\zeta\half}
\label{fnlofts2} \\
\tnl &=& 
   \ts \( \frac{
\calp_{\zetas}
}{ 
{\calp_\zeta} 
}\)^2 
\frac1{\calp_\zeta}
\label{tnlofts2}
.
\eea

\subsubsection{The general case}

In general, \eq{70} will contain higher terms
\be
V(\sigma) = \frac12m_*^2\sigma^2 + \lambda\sigma^4 + \sum_{d>4}
\lambda_d \frac{\sigma^d}{\mpl^{d-4} }
\label{vextra}
.
\ee
They will not affect the Bunch-Davies result  provided that
\bea
\lambda &\ll& \ts^2  \label{lamcon}  \\
\lambda_d \( \frac{\sigma}{\mpl} \)^{d-4} &\ll & \ts^2
\label{lamdcon}
.
\eea

For an  arbitrary potential $V(\sigma)$, assuming still that 
inflation with practically constant $H$ lasts for long enough, 
 the
  probability 
of finding $\sigma $ in a given interval is   \cite{stochastic,sy}
\be
P(\sigma) d\sigma  \propto  \exp(-8\pi^2 V /3H_*^4) d\sigma 
\label{vgauss}
.
\ee
The  potential
is $V(\sigma)$ is evaluated with all other relevant fields fixed,
with the convention that its minimum vanishes.

 There are two 
simple cases.
If $V(\sigma)$ increases until it becomes at least of  order $H_*^4$,
the  probability distribution is more or less flat out to
a value $\sigma\ma$, such that 
$V\sim 3H_*^4/8\pi^2$.\footnote
{This differs from the estimate of the typical value given in
\cite{dlnr,dllr,dllr2}.} For the Bunch-Davies case one can take $\sigma\ma^2$
to be the variance $(H_*/2\pi)^2/\ts$ of the gaussian probability distribution.
In this case the tilt is positive and it can be strong, corresponding to
the lightness condition $\ts\ll 1$ being only marginally satisfied.

Instead,   $\sigma$ might be  a PNGB with the potential
\be
V(\sigma) =  \Lambda^4 \cos^2 (\pi\sigma/f)
\label{pngb}
.
\ee
If $\Lambda \ll H_*$ the  probability distribution for $\sigma $
is  extremely flat within the fundamental interval 
$0<\sigma<f$. The  effective mass at the maximum and minimum of 
the potential is $m_*=\pi \Lambda^2/f$ giving the maximum tilt as
\be
|\ts| \simeq  \frac{\Lambda^2}{H_*^2} \frac{\Lambda^2}{f^2}
\label{tspngb}
.
\ee
To have a reasonable probability for being at the maximum 
requires $\Lambda\ll H_*$, and to have weak self-coupling so that the 
semi-classical theory used here makes sense requires $\Lambda\ll f$.
 Judging by this example, strong negative
tilt looks unlikely and is not considered in the present paper.

On the other hand, slight negative tilt consistent with observation is
possible. The  probabilities for being at the maximum and minimum of the
potential are related by
\be
\frac{P\sub{max}}{P\sub{min}} = \exp\(-\frac{8\pi^2\Lambda^4}{3H_*^4} \)
,
\ee
and one can have say $\ts=-0.05$ while keeping this ratio not too far
below 1.

It must be emphasized that the probability distribution 
 \eq{vgauss}  is attained only if  the variation of $H$ is negligible,
on the timescale for the rolling down of $\sigma$ towards its zero value.
Whatever it is, the late-time 
probability distribution is not relevant for the inflaton in a 
single-component inflation model, since its value a given number of 
$e$-folds before the end of inflation is obtained by integrating
the trajectory $3H\dot\phi=-V'$.

If eternal inflation takes place around some maximum of the potential,
$H$  will be practically constant during the eternal inflation and
all light fields will attain the probability
distribution \eqreff{vgauss} except for the one driving eternal inflation.
When eternal inflation ends and slow-roll inflation begins,
 the fields orthogonal to the 
inflationary trajectory will have this probability distribution.
In a single-component inflation model it
 will still apply when the observable Universe leaves the horizon,
if the value of $V$ 
then is not much lower than it was during eternal inflation.
That may be the case if  the potential has a suitable maximum,
which  is more likely than one might think
\cite{blhilltop}. The  distribution \eqreff{vgauss}
might also be attained \cite{lm2}
if eternal inflation 
occurs high up on  the chaotic inflation potential $V\propto \phi^2$.
A further possibility, not yet investigated, is that 
the  distribution \eqreff{vgauss} is attained if eternal inflation occurs
at a {\em maximum} of the potential  with high  $V$;
 possibly well-motivated \cite{extranat,knp}
realizations  of that case would be Natural 
Inflation \cite{natural} along with  its hybrid \cite{ewanmulti,pseudonat}
and multi-component \cite{nflation} generalizations.

\section{The  curvaton model}
\label{five}

\subsection{The setup}
The  curvaton model \cite{mollerach,lm,curvaton,moroi}
(see also \cite{es}) is a particular realization of \eq{95}.
The  curvaton field $\sigma$ at some stage is  oscillating  harmonically
about  $\sigma=0$ under the influence of a quadratic  potential 
$V=\frac12 m^2\phi^2$, with energy density $\rho_\sigma\propto a^{-3}$.
This stage begins at roughly the epoch 
\be
H \sim m
\label{oscepoch}
.
\ee
This  mass $m$  in these equations is  taken to be the true vacuum mass, 
 the idea being
that the effect of other fields on $V(\sigma)$ will have become negligible
by this time.  That  will be more or less  true
 if the effective mass up to that
time has been  $\lsim H(t)$. 

When  the harmonic oscillation begins, 
 $\rho_\sigma$ is supposed to be negligible compared with the total
$\rho=\rho_\sigma+ \rhoin$. The component $\rhoin$ (defined as  the
difference between $\rho$ and $\rhos$)  
roughly speaking  originates from the decay of the inflaton but 
there is no need of  that interpretation.
The curvaton contribution to the curvature perturbation is  at this stage
supposed be negligible.

Eventually the harmonic oscillation decays. 
During  at least some of the oscillation era,
$\rho\sub{inf}$ is  supposed to be radiation-dominated 
 so that $\rhos/\rhoin$ 
 grows and with it
$\zeta$.\footnote
{Equations are derived on the assumption that these quantities
are initially negligible. Presumably those same equations will provide
a crude approximation even if the growth is negligible, due either to 
the curvaton decaying promptly \cite{yeinzonprompt,frv} or to $\rho\sub{inf}$
containing a negligible radiation component.}

Originally $\rhoin$ was supposed to be radiation-dominated
during the whole oscillation era, but the
 model is not essentially altered if $\rho\sub{inf}$
 contains a significant contribution from matter.
This matter might be  the homogeneously oscillating
inflaton  field which decays only after the onset of the curvaton
oscillation
\cite{dlnr,dlr,dllr2},  
non-relativistic curvaton particles \cite{lm2},  other
 non-relativistic particles which decay before the curvaton
or any combination of these.

The lightness of the curvaton field can be ensured
 by taking it to be a PNGB with the potential
\eqreff{pngb}. This mechanism can work whether or not there is supersymmetry,
and is easier to implement for the curvaton  than for the inflaton
\cite{curvaton,dlnr,cdl}.

Several 
 curvaton candidates exist which were proposed already for other reasons.
Using such a candidate, one might  connect the origin of the curvature
perturbation  with particle physics
beyond the Standard Model,  or even with observations at
colliders and detectors. 
 Among the candidates are a right-handed
sneutrino \cite{serendip,rhsnu,postma,hkmt,mt}, a modulus 
\cite{moroi,hkmt} (which might be a string axion \cite{dlnr,stringaxion}),
 a  Peccei-Quinn field \cite{dllr,pqcurvaton} 
and an MSSM  flat direction \cite{flatdirection,hkmt}. The 
 right-handed sneutrino possibility was 
actually discovered serendipitously \cite{serendip}
by authors who were unaware of the 
curvaton model, which shows  that
 the curvaton model is not particularly
contrived. 

\subsection{The master formula}

In this subsection  the basic approach is that of
 \cite{curvaton,luw}, which works with the 
the first-order perturbation theory
 expression \eqreff{zetaofdrho}.
This is applied after the onset of the harmonic oscillation of the 
curvaton, when the cosmic fluid has two components. 
%One, the oscillating
%curvaton, has pressure $P_\sigma =0$. The other component
%comes from inflaton decay and preheating.
%The essential assumption of the curvaton model is that the curvaton
%component  gives a negligible contribution to $\zeta$
%at the onset of the oscillation, its contribution growing subsequently
%because the inflaton contribution is radiation-dominated for at least part
%of the time. 
The final value of 
 $\zeta$ is taken to be the one evaluated 
just before the curvaton decays, which  is taken to occur instantaneously
on a slice of uniform energy density,
 at an  epoch $H\sim \Gamma$ where $\Gamma$ is the decay width. 

Keeping this  basic  setup, 
 the treatment of \cite{curvaton,luw} will be generalized to allow
for several possible effects.\footnote
{Each effect was considered  before, usually without 
any of the others.
Strong tilt and non-gaussianity were
 considered in \cite{lm,lm2}. The 
possible contribution of
curvaton particles to $\rhoin$ was   taken into account qualitatively in
 \cite{lm2}.
 The Bunch-Davies case
was considered  in \cite{postma,lm,lm2}. Curvaton
 evolution  
 was partially taken into account  in \cite{dllr2,luw,mylow}
(see also  \cite{lr} for a treatment using the $\delta N$ formalism)).
The possible  contribution of $\zetag$
was taken into account in \cite{moroi,dlnr,lv}.
All of this is at  first order. The calculation to second order in 
cosmological perturbation theory was
done in \cite{bmssecondorder,mwsecondorder} 
(re-derived in \cite{lr} using the $\delta N$ formalism.)
 Also,  the sudden-decay
approximation was removed  in \cite{mwu}, 
at first order only.}
The inflaton component $\rhoin$ is not required to be purely radiation.
The contribution of the inflaton perturbation 
to $\zeta$ is not required to be negligible.
The  tilt $\ts$  is taken into account.
Attention will focus on constant tilt which is either  small 
($|\ts|\lsim 10^{-2}$, or else
 large and positive, which is allowed if the curvaton contribution 
to $\zeta$ is sub-dominant. In that case strong non-gaussianity will also be
allowed.

Evolution of the curvaton field after the end of inflation will
be allowed. One possibility \cite{dllr2}
for such evolution is the large
effective mass-squared
 $V_{\sigma\sigma}\sim \pm H^2(t)$ predicted by supergravity 
for a generic field during matter domination
(though not   \cite{ltakeo} during radiation domination). 
A more drastic possibility is for 
$\sigma$ to be  a PNGB corresponding to the angular part of a complex field,
whose radial part varies strongly \cite{mylow,cdl,dlr,dimlaz}. 

%Finally, the component the inflaton component of the fluid 
%is not required to be radiation-dominated throughout. In particular,
%it is allowed to contain non-relativistic curvaton particles, produced at for
%instance the decay of the inflaton \cite{lm2}. 

In the presence of evolution, 
the oscillation may initially  be anharmonic, but after a few Hubble times the
amplitude presumably will have decreased sufficiently that the oscillation is
harmonic, making   \eq{oscepoch}
 an adequate approximation. (A detailed discussion for the analogous
axion case is given in \cite{myaxion}.)

The evolution is  given by
\be
\ddot \sigma + 3H(t)\dot\sigma + V_\sigma = 0 
\label{sigmaeq}
.
\ee
Since the inflaton perturbation just corresponds to
a shift in time, the `separate universes' are practically identical
until after the onset of the oscillation, and 
 \eq{sigmaeq}
 holds locally at each position. Let us define
the  amplitude $\so(\bfx)$ at the start of
the harmonic oscillation  on a spacetime slice of uniform 
energy density. Then 
$\so(\sigmae(\bfx))$ is a function only of $\sigmae$, 
 the $\bfx$-dependence coming purely from the fact that 
$\sigmae(\bfx)$ is not defined on such a slice.
If $V$ is quadratic (with a constant or slowly-varying mass)
$\so(\sigmae)$ is practically linear.

Knowing $\so(\bfx)$ we can calculate\footnote
{Recall (footnote 2) 
that there is no need to demand $\overline{\delta\rhos}=0$.}
\bea
\rhos(\bfx) &=& \frac{m^2}2 \( \so + \delta\so(\bfx) \)^2 \\
\bar\rho_\sigma &=& \frac{m^2}2 \overline{\so^2} \\
\overline{\so^2} &=& \so^2 + \vev{\delta\so^2} \\
\delta\rhos &=&  \frac{m^2}{2} \( 2\so \delta\so + \delta\so^2 \) 
.
\eea

To go further we expand $\so(\sigmae)$ to second order in $\delta\sigmae$
giving \cite{mylow}
\be
\delta\so(\sigmae(\bfx)) =  \so' \delta\sigmae(\bfx) +
\frac12 \so'' \delta\sigmae^2(\bfx)
\label{osexp}
,
\ee
and
\bea
\delta\rhos  &=& \frac{m^2 \so^2} 2 \[ 2q\frac{\dss}{\sigmae}
+ u \( \frac{ q \dss}{\sigmae} \)^2 \]
\label{delrhos}  \\
\bar\rho_\sigma &=&  \frac{m^2p}{2} \so^2 \\
q &\equiv&  \sigmae\so'/\so \\
u &\equiv& 1 + \so \so''/ \so'^2 \\
p &\equiv& \frac{ \overline{\so^2} } {\so^2} =
1 +  uq^2 \frac{ \vev{\delta\sigmae^2}  } { \sigmae^2 } 
\label{pexp}
.
\eea
If $\delta\sigma$ has negligible evolution,  or if $\sigma\os(\sigmae)$
is linear corresponding to a quadratic potential,
then
$q=u=1$. It has been shown \cite{en} that even slight anharmonicity
could in certain cases give $|u|\ll 1$, and as we have seen strong
evolution is also possible making both $q$ and $u$ very different from 1.

Except in Section \ref{s:bd}, all  calculations will be done with 
the   minimal box size, corresponding to $\ln(LH_0)$ not too far above 1.
We need $\vev{\delta\sigmae^2}$;
\bea
\vev{\delta\sigmae^2}
 &\simeq &  \( \frac{H_*}{2\pi}\)^2
\int^{k\sub e}_{L\mone}   \frac{dk}{k} \( \frac k{k\sub e} \)^{\ts} \\
&=&   \( \frac{H_*}{2\pi}\)^2 y( L k\sub e)
.
\eea
This gives
\be
p=1 + uq^2 \( \frac{H_*}{2\pi\sigma} \)^2 y(L k\sub e) e^{N_L\ts}
\label{bestpeq}
.
\ee
%where $\sigma$ is the average field in the observable Universe when it
%leaves the horizon.
with 
\be
y \simeq  \min  \left\{ \begin{array}{l} N   \\
                                    1/\ts 
                \end{array}  \right. \label{ychoices}
\ee
A crude but usually adequate approximation is $ \vev{\delta\sigmae^2} 
\simeq H^2$.
(See \cite{myaxion} for a similar estimate of the axion perturbation.)

We have been evaluating $\rhos$ and its perturbation at the beginning of 
the oscillation, on a slice where $\rho=\rhoin$ is uniform. 
 The curvature perturbation is given by \eq{zetamulti}
in terms of the perturbations of the two fluids evaluated on the flat slicing.
The curvaton perturbation on the uniform-density 
slicing is
\be
 \frac{\delta\rho_\sigma }{\overline \rho_\sigma} =
\( \frac{\delta\rho_\sigma}{\overline\rho_\sigma} -
\frac{\delta\rhoin}{\rhoin+P\sub{inf}}  \)\sub{flat}
.
\ee
Each term in this expression is time-independent  \cite{luw}.
 (Remember that the two-fluid
description is only valid during the harmonic oscillation.)
Using it,   \eq{zetamulti} evaluated
 just before the curvaton decay gives\footnote
{According to the definitions made in this paper, $\hat\zeta\sub{inf}
=\zeta\sub{inf}$, but $\hat\zeta_\sigma\neq \zeta_\sigma$.}
\bea
\zeta&=& \zetag + \zetas  \\
\zetas &=& f  \frac{\delta\rho_\sigma }{\bar\rho_\sigma}
 \\
3 f 
 &=&  \frac{\bar\rho_\sigma}{\rho + P} 
=\frac{\bar\rho_\sigma}{\bar\rho_\sigma + \rhoin + P\sub{inf}}
\simeq  \frac{\bar\rho_\sigma}{\rho} \equiv \oms
.
\eea
The   approximation is adequate,  because the 
`sudden-decay' approximation  has generally a significant error
\cite{mwu} in the regime $\oms<1$.

The value of $\oms$ is to be calculated at the decay epoch $H\sim \Gamma$.
It is sometimes convenient
to  write
\be
\Gamma = \gamma m^3/\mpl^2
.
\ee
Then $\gamma \sim 10^{-2}$ corresponds to gravitational-strength decay
\cite{thermal2} and one expects $\gamma\gsim 10^{-2}$.

Suppose first that there is continuous
 radiation domination during the oscillation.
  If $\oms \ll 1$,  \cite{curvaton}
\be
\oms \simeq
 \frac{\overline{\so^2} }{\mpl m \gamma\half}
\label{ommax1}
.
\ee
An approximation valid for any $\oms$ is therefore
\bea
\oms &\simeq& \frac {\overline{\so^2} }{\overline{\so^2} + C^2 }
\label{omparam}  \\
C^2 &= &  \mpl m \gamma\half  = \mpl^2 \sqrt\frac \Gamma m
\label{csquared}
.
\eea

Requiring a decay rate of at least gravitational strength,
the  first equality implies 
\be
C^2\gsim 10\mone \mpl m
\label{ommax0}
.
\ee
Requiring that the decay takes place before the onset of 
nucleosynthesis, corresponding
to $\rho\quarter > 1\MeV$, the second equality implies
\be
C^2 \gsim 10^{-21} \mpl^{5/2}/m\half
\label{ommax}
.
\ee
These bounds cross at
$m\sim 10^4 \GeV$, implying $C \gsim 10^{11}\GeV$.
It will be important later that $C$ might be either bigger or smaller
than $H$.

If $\rhoin$ has a matter component $C$ is bigger.
In particular, a contribution of 
 curvaton particles, denoted
by $\rho\sub c$, gives
%\be
%\frac{\rhos}{\rho\sub c} \sim \frac{\overline\so^2} { 
%\Omega\sub c \mpl^2
%}
%,
%\ee 
\be
C^2 \simeq  \mpl m \gamma\half + \mpl^2 \Omega \sub c
,
\ee
where $\Omega\sub c$ is evaluated at the onset of the oscillation.
(The useful parameterization \eqreff{omparam} was first given in \cite{lm2},
keeping just the contribution of curvaton particles.)

Using these equations
 we arrive at the master formula;
\bea
\zetas  &\simeq & \frac{2\oms}{3 p}
\[ q\frac{\dss}{\sigmae}  + \frac u 2  \(q\frac{\dss}{\sigmae}\)^2 \]
\label{zetafinal}
.
\eea
After adjusting the normalization of $\delta\sigma$ this has the form
 \eq{withlin}. Then the correlators
are given by \eqst{speczero}{tsigs}.

\subsection{Special cases}

\label{s:special}

If we consider a single scale,
\eqst{pzeta}{tnlofy} apply, and if in addition this scale is taken
to be $k\sim H_0(\sim L\mone)$ we can write things in terms of $\sigma$;
\bea
\zetas  &\simeq & \frac{2\oms}{3 p}
\[ q\frac{\delta\sigma}{\sigma}  + \frac u 2  
\(q\frac{\delta\sigma}{\sigma}\)^2 \]
\label{zetafinal2} 
,
\eea
with $p$ given by \eq{bestpeq}. With that understanding let us evaluate
the correlators in some special cases.

Let us  assume that  $\zetas$ is  dominated by the linear term, corresponding
to 
\be
4 \frac{q^2 u^2}{\sigma ^2} \( \frac{H}{2\pi} \)^2
 \ll 1
\label{lincon}
.
\ee
Then
\bea
\calp_{\zetas}\half &=&  \frac{2\oms  q}{3 p}
\frac{H}{2\pi\sigma }    \\
\frac35 \fnl &=& \frac{3pu}{4\oms}
\( \frac{\calp_{\zetas}} {\calpz} \)^2 \\
\tnl &=& (36/25) f\sub{NL}^2 \( \frac{\calp_{\zetas}} {\calpz} \)^3
.
\eea

In this case $\zetas$ can be the dominant contribution, demanding
$e^{N\ts}\sim 1$. Then, unless $u$ is extremely small, $p\simeq 1$
leading to
\bea
\calp_{\zeta}\half &=&  \frac{2\oms  q}{3 }
\frac{H}{2\pi\sigma } \label{th0}   \\
\frac35 \fnl &=& \frac{3u}{4\oms}
\label{th1}
 \\
\tnl &=& (36/25)f\sub{NL}^2
\label{th2}
.
\eea

For  $q=u=1$, corresponding to
negligible evolution or evolution under a
quadratic potential,  \eqs{th0}{th1} reduce to the
 standard result \cite{luw};
\bea
\calp_{\zeta}\half & \simeq  &  \frac{\oms H}{3 \pi \sigma}
\label{pzs}  \\
\frac35 \fnl &=& \frac34 \frac 1 {\oms } 
\label{pzs2}
,
\eea
the bound on $\fnl$ requiring $\oms \gsim 10\mtwo$.

Supposing further that the evolution actually is negligible,
\be
\oms=\frac{\sigma^2}{\sigma^2 + C^2}
.
\ee
This is the simplest version of the curvaton model. It gives
 $H\gsim \calpz\half  C$, and then \eq{ommax} gives \cite{mylow} the bound
\be
H\gsim 10^7\GeV
\label{myhbound}
.
\ee

%Now suppose that \eq{lincon} is reversed, so that
% $\zetas$ is dominated by the quadratic term. Then  $\zetag$
%must be dominant, and 
%\be
%\calp_{\zetas}\half   \simeq 
%\frac{\oms  |u| q^2}{\overline{\sigma ^2}  }
%\( \frac{H}{2\pi} \)^2 
%.
%\ee
%As we saw, this will be detectable if it is of order $10\mone \calpz\half$.

\subsection{The case $\oms = 1$}

In the limiting case where $\oms$ is indistinguishable from 1 there is no
need of the sudden-decay approximation. Before  curvaton decay is appreciable,
the curvaton-dominated cosmic fluid has $P=0$ making  $\zeta$, and hence
$\zetas$,  a constant. The local value of $\sigma$ provides the initial
condition for the evolution of the separate universes, making them
 identical. As a result, $\zeta$ remains constant throughout and after
the curvaton decay process.

To evaluate the non-gaussianity in this case
 one can use the $\delta N$ formalism,
which for the curvaton model  is equivalent to using second-order 
cosmological perturbation theory.
 Adopting the 
small-tilt and $p=1$ assumptions,
the calculation described in \cite{lr} applies so that
\be
\frac35 \fnl =  \frac34 u - \frac32 
.
\ee
With negligible evolution or a purely quadratic potential, 
$\fnl= -\frac54$. It would be interesting to know if such a small
value will ever be observable.

\subsection{Induced isocurvature perturbations}

The status of 
isocurvature perturbations in the curvaton model is considered
elsewhere \cite{luw,lwcdm} on the assumption that $\rhoin$ contains no curvaton
particles. Let us briefly reconsider the situation when that assumption is
relaxed. 

 As defined
by astronomers, an isocurvature perturbation $S$
may be present  in any or all of the
baryon, cold dark matter or neutrino components of the cosmic fluid
when cosmological scales first approach the horizon,  
 being the 
fractional perturbation
in the relevant number density on 
 a slice of uniform energy density.
 Given the  separate
universe assumption,  the 
 inflaton perturbation $\delta\phi$ cannot generate an isocurvature
perturbation since it just corresponds to a shift back and forth along
the inflaton trajectory. Any  orthogonal light field $\sigma_i$ might
create an isocurvature perturbation, and the same field might give the 
dominant contribution to the curvature perturbation so that the two
perturbations would be fully correlated.
This has
 been called a 
residual isocurvature perturbation \cite{luw,lwcdm,lwcons}.

In the  curvaton model, a residual isocurvature perturbation 
obviously cannot be created after the curvaton decays. 
If the cdm or baryon number is 
is created {\em by} the curvaton decay and $\rhoin$ contains no
curvaton particles, the argument of \cite{luw} gives
 a residual isocurvature
perturbation  $S\simeq - 3(1-\oms) \zeta$.
This is  viable only if $\oms$ is close to 1.
 Repeating the argument of \cite{luw}
for the case that $\rhoin$ contains 
curvaton particles, one easily  sees that they 
should be discounted when evaluating $\oms$ 
in the expression for $S$, allowing a true
$\oms\ll 1$. 

Finally, suppose that 
 the cdm or baryon number
 is created before curvaton decay. Then, 
whether or not $\rhoin$ contains curvaton particles,   
 the argument of  \cite{luw} gives a residual isocurvature
perturbation $S/\zeta\simeq -3(1-\Omega_\sigma\su{crea})$ 
where the superscript 
denotes the epoch of creation.

 The above formulas apply also to the fractional isocurvature perturbation
in the lepton number density, from which the neutrino isocurvature perturbation
can be calculated \cite{luw}. It is not out of the question to generate the
big lepton number density,  that is needed to give  a significant
 neutrino isocurvature perturbation \cite{withjohn}.

\subsection{The Bunch-Davies case}

\label{s:bd}

So far $\sigma$ is unspecified. This is the unperturbed value of the curvaton
field when the minimal box leaves the horizon. Now we consider the case
that $\sigma^2$ is equal to the variance $(H/2\pi)^2/\ts$
of the 
Bunch-Davies distribution. As we saw earlier, it becomes simpler in that case
to use a maximal box, such that $\sigma\sub e=0$ and
\be
 \overline{\sigma\sub e^2} = \vev{\delta\sigma\sub e^2} = \( \frac{H_*}{2\pi}
\)^2 \frac1{\ts}
.
\ee
Then the master formula can be written 
\bea
\zetas &=& \frac13 \oms u q^2 \hat p \frac{\delta\sigma\sub e^2}
{ \vev{\delta\sigma\sub e^2} } \label{161x} \\
\hat p &\equiv & \frac{
\vev{\delta\sigma\sub e^2} 
}{
\overline{\sigma\os^2}
}
= \frac{
 \overline{\sigma\sub e^2}
}{
\overline{\sigma\os^2}
}
.
\eea
In the following I set equal to 1 the  evolution factor
 $qu\hat p$.

After adjusting the normalization of $\delta \sigma$,
 \eq{psofts2} gives
\be
\calp_{\zetas}\half = \frac23 \oms t_\sigma\half e^{-N\ts} \( \frac k{H_0} 
\)^{\ts} \label{supz}
.
\ee
If $\calpz= \calp_{\zetas}$, observation requires $N\ts\lsim 1$
and \eqst{psofts2}{tnlofts2} give
\bea
\calp_\zeta\half &\simeq & \frac23 \oms \sqrt{\ts}
 \label{bdpzeta} \label{166}\\
\frac35 \fnl &\simeq  & \frac3{4\oms} \label{bdfnl}
\label{167} \\
\tnl &=& (36/25) \fnl^2 \label{166a}
.
\eea
In this case 
$\ts \sim 10^{-9}/f\sub{NL}^2$, which has to be very small indeed as
 was  first noticed  by Postma \cite{postma}.

If $\calp_{\zetas}\ll \calpz$,
strong tilt is allowed. Then \eqs{fnlofts2}{tnlofts2} give
\bea
\frac35\fnl &=& 4 \ts^2 \(\frac{\oms}{3} \)^3 
e^{-3N\ts} \( \frac k{H_0} \)^{3\ts} \calpz^{-2} \\
\tnl &=& 
16 \ts^3 \(\frac{\oms}{3} \)^4 
e^{-4N\ts} \( \frac k{H_0} \)^{4\ts} \calpz^{-3}
.
\eea
The present bound  $\fnl<121$ requires $\oms e^{-N\ts}< 9\times 10^{-6}$
and the present bound  $\tnl<10^4$ requires  the stronger bound
$\oms e^{-N\ts} < 2\times 10^{-6}$. 
These bounds are an extension of  \eq{rngexp}, derived now
for a maximal box. As in that case, they apply only on large cosmological
scales, allowing a large curvature perturbation n much smaller scales.

To summarize, we find in the Bunch-Davies case
 strong non-gaussianity on 
large scales ($k\sim H_0$) provided that
 $N\ts \gsim 1$. This is because 
 a typical region of size $H_0\mone$ then becomes
very inhomogeneous by the end of relevant inflation.
The non-gaussianity is reduced as the scale is decreased,
and (with $\oms=1$) is small  on the scale leaving the horizon
at the end of relevant inflation  precisely because a typical region 
of size $k\sub e\mone$ is still quite homogeneous.

Although the maximal box provides the neatest result, it is interesting
also to see what happens with a minimal box. 
For a box of any size, \eq{161x} becomes
\be
\zetas = \frac13 \oms u q^2 \hat p  \vev{\delta\sigma\sub e^2}\mone
\( 2\sigma \sub e \delta\sigma \sub e + \delta\sigma\sub e^2 \)
.
\ee
Setting $u=q=\hat p=1$ and adjusting for the normalization of $\delta \sigma$,
we can calculate the correlators of $\zetas$ from 
\eqss{36a}{fnlofy1}{tnlofy1} and then take their expectation values within a
 maximal box, to find
\bea
\vev{\calp_{\zetas}} &=& 4 \ts \(\frac{\oms}{3} \)^2  e^{-2N\ts} 
\( \frac k{H_0} \)^{\ts} f \\
\frac35\vev{\fnl} &=& 4  \ts^2 \(\frac{\oms}{3} \)^3 e^{-3N\ts} 
\( \frac k{H_0} \)^{2\ts}\calpz\mtwo f \\
\vev{\tnl} &=& 16 \ts^3 \(\frac{\oms}{3} \)^4  e^{-4N\ts} 
\( \frac k{H_0} \)^{3\ts}\calpz\mthree f \\
f &\equiv&  (LH_0)^{-\ts} + y(kL)  \ts \( \frac k{H_0} \)^{\ts}  
.
\eea
In accordance with the discussion of Section \ref{two} these expressions 
are independent of the box size. But the split into the linear plus quadratic
term depends on the box size.
With a  maximal box the linear term vanishes,  even in gaussian
regime $N\ts\lsim 1$. 
With  a minimal box the linear term  dominates  on large scales,
even in the non-gaussian regime.

With a minimal box, negligible tilt and $\zeta=\zetas$, \eq{pzs}
applies. Inserting the Bunch-Davies expectation value for $\sigma^2$
then reproduces \eqst{166}{166a}.
If instead we  consider the  PNGB case, assuming  $\Lambda\ll H_*$ so that 
$\sigma$ has a flat distribution up to $\sigma\max=f$, 
 \eq{166} becomes 
\be
\calpz\half \gsim \oms \frac{ H_*} \Lambda \frac \Lambda f
\gg \oms \frac \Lambda f
.
\ee
Using \eq{tspngb} we have again $\oms\sqrt{\ts} < 1$, making $\ts$
indistinguishable from 1.

\subsection{Cosmic uncertainty}

 To make
contact with previous work, I assume in this subsection 
that the curvaton field has negligible spectral tilt, and has
negligible evolution before the oscillation starts.
Also I consider $Q\equiv (2/5) \calpz\half$
whose observed value is $2\times 10^{-5}$.

We are not going to be concerned with precise values, but in this context
we do not want to exclude the  case $\sigma^2 \lsim H^2$.
To handle it we can replace $\sigma$ in \eq{pzs} by $\sqrt{\sigma^2+H^2}$, 
leading to 
\bea
Q &\sim &  \frac{
\sqrt{\sigma^2 + H^2} }
{\sigma^2 + H^2 + C^2}
H \sim  \oms \frac{H}{\sqrt{\sigma^2 + H^2 }} \label{qexp}  \\ 
f\sub{NL}\mone &\sim & \oms \sim 
 \frac{\sigma^2 + H^2}{\sigma^2 + H^2 + C^2} \label{fnlexp}
\eea
{}In  this expression, $H$ is evaluated during inflation.

In the Bunch-Davies case, the probability distribution for $\sigma$
is gaussian, making it more or less flat up to a maximum value
of order the variance; $\smax \sim H/\sqrt{\ts}\gg H$.
In the PNGB case (\eq{pngb} with $\Lambda\ll H$) the 
the probability distribution is almost perfectly flat, up to a smaller
maximum which could be much less than $H$.
Using the  flat distribution,  one can work out the
non-flat distribution for the correlators. 
To discuss this, I take 
the  tilt $\ts$  to have a small fixed value,
consistent with  observation.

There are three parameters $C$, $H$ and $\sigma$ and I will divide the 
parameter space into regions, separated by strong inequalities to allow
simple estimates.

Consider first the case $C\ll H$. In the regime  $H \ll \sigma \ll \sigma\ma$
we have 
$Q\simeq H/\sigma$,  with $\oms$ very close to $1$ and $\fnl=-5/4$.
In the regime $\sigma \ll H$, the quadratic term dominates
$\zeta$ making $Q\sim 1$  in contradiction with observation.
Given the flat distribution for $\sigma$, the  lengths $\sigma\ma-H$
and $H$ over which $\sigma$ runs for the two cases give their
relative probabilities for a randomly-located region.

Now consider the case $C\gg H$. 
There are  three regimes;
\begin{enumerate}
\item The nearly gaussian  regime $C\ll \sigma < \sigma\sub{max}$.
Here $Q\sim H/\sigma$ and $\oms\simeq 1$  making $\fnl\simeq -5/4$.
\item The strongly 
non-gaussian regime $H \ll \sigma \ll C$. Here $Q\sim H\sigma/C^2$
and $f\sub{NL}\mone \sim \oms\sim \sigma^2/C^2 \ll 1$. The lower part of this
range is forbidden by observation.
\item The  regime $\sigma \ll H$. Here the quadratic term
of \eq{zetafinal} dominates leading to $Q\sim H^2/C^2$,  and to
$\fnl\sim 1/Q$ in contradiction with  observation.
\end{enumerate}

The relative probabilities, that  a given
 region corresponds to one or other of these cases,  are proportional to the 
intervals $\Delta \sigma$ given at the beginning of each item.
Within each case, the probability distribution for 
 $Q$ is 
\be
dP = d\sigma = \frac{d\sigma}{dQ} dQ
\label{dpofdq}
.
\ee

\subsection{Anthropic considerations}
\label{s:an}

So far we did not take into account anthropic considerations. They have gained
 force recently, since it appears that string theory allows a very large
number of field theory lagrangians.
 Both the idea and the methodology of anthropic
arguments are very controversial, as emphasized for instance in
\cite{lm2}, but let us proceed. 

Anthropic arguments  suggest \cite{tr} that $Q$ has to be in a 
  range  $Q\mi < Q < Q\ma$
corresponding roughly to 
 $10^{-6} \lsim Q \lsim 10^{-4}$.
If the cosmological constant is taken to be fixed the probability distribution
within this range is more or less flat. But Weinberg argued \cite{weinberg}
 that the cosmological constant itself should
be regarded as having a flat probability distribution, since there appears
 to be no theoretical argument that would give a definite value, in particular
zero.
 He showed, before the data demanded it,
 that anthropic arguments suggest  a value
appreciably different from zero. As summarized in \cite{gv}, 
subsequent studies 
have shown the preferred value to be compatible with
observation. Accepting this viewpoint for the cosmological constant, it has 
been argued \cite{glv} that 
the probability distribution of $Q$ within the above range is 
\be
dP \propto  Q^3 dP\sub{prior} 
\label{gvexp}
,
\ee
where the prior $dP\sub{prior}$  is the 
probability distribution if anthropic considerations are ignored.
I will use this estimate in the following discussion, without trying
to take on board the impact of some more recent work \cite{tarw}.

At this point, one may wonder why the focus is exclusively on the
overall normalization $Q^2$ of the spectrum. What about the spectral 
tilt, and the measures $\fnl$ and $\tnl$ of non-gaussianity?
The original arguments of Harrison \cite{harrison} 
and Zeldovich \cite{zeldovich} for tilt in the range
$|\ts|\lsim 1$ may be regarded as anthropic, but we are now dealing with
an observational bound more like $|\ts|<0.01$.
It  seems clear that no
anthropic consideration will directly 
produce this result, and  there was no
objection to values
 $|\ts|\sim 1$ before the cmb anisotropy ruled them out.
Coming to non-gaussianity, it is again hard to see how
 anthropic consideration will directly constraint it,
and  there was no  objection even to the extreme case
$\fnl\sim \calpz\mhalf$ until it was ruled out by observation.
{}From the anthropic viewpoint, 
 the small tilt and non-gaussianity 
presumably are produced accidentally, by anthropic constraints
on other parameters including $Q$.

If $Q$ is generated by the inflaton perturbation in a single-component
model, it depends almost entirely
 on some parameters in the field-theory lagrangian.
(There is some  dependence on the post-inflationary history via
 $N$ but it would take a big variation of that history
to have much effect on $N$.) Then the tilt also depends only on the field
theory  parameters,
while the non-gaussianity is automatically negligible.
If  the field theory parameters were 
  taken  to be fixed,  the prior would be    a delta
function and there would be  no room for anthropic arguments. 

In contrast, curvaton-type models depend also on the background values
of one or more fields, and if inflation begins  early enough one has no
option but to consider their variation within the very large and smooth
inflated patch that we  occupy. This was pointed out some time
ago in \cite{cdl,dlnr} for the PNBG case, and has been discussed more recently 
in \cite{lm2} for the Bunch-Davies case. The point here is that 
 anthropic considerations concerning $Q$ may
demand,  or anyhow favour, a value $\sigma$ far below $\smax$.

A precisely similar situation exists with respect to the nature of the
CDM.
If it consists of  neutralinos, or of  axions created by the oscillation
of cosmic strings, the CDM density is given in terms of parameters
of the lagrangian. If instead it is the oscillation of  a nearly  homogeneous
 axion field
 which existed during inflation, the CDM
density varies with our location within the inflated patch (and so does
the CDM isocurvature perturbation which is inevitable in that case
\cite{myaxion}). Linde \cite{lindeaxion}
 provided the first concrete realization of anthropic
ideas, when he pointed out that we might
need to live in a place where the
axion density is untypically small.
The anthropic probability for the  axion density has recently 
been investigated \cite{tarw}.\footnote
{In this case the axion density is proportional to
$(\sigma^2  + H^2)$ where $\sigma$ is the average axion field in our
part of the Universe and $H^2$ comes from the long wavelength fluctuations
\cite{myaxion}. The authors of \cite{tarw} drop the $H^2$ term, 
which might possibly affect their results.}

Now I analyse the situation for the 
 actual curvaton model,
 generalizing
two recent discussions \cite{gv,lm2}.
The spectrum $Q^2$ is  given by  \eq{qexp}, and we are assuming that the
probability distribution is flat within a range $0<\sigma<\sigma\ma$.
There is also the the anthropic constraint $Q\mi < Q < Q\ma$. These
 inequalities
define a rectangle in the $Q$-$\sigma$ plane, and we can only use the 
part of the curve \eqreff{qexp} that lies within this rectangle.

The location of the rectangle relative to the curve depends on the parameters
$C$ and $H$ which define the curve, and on $\sigma\ma$ whose value was 
discussed earlier. In this 3-parameter space, there will be an unviable
 regime 
where no part of the curve lies within the rectangle. 
In the opposite case, part of the curve is within the allowed rectangle,
and putting  \eq{dpofdq} into \eq{gvexp} gives the probability distribution
for $Q$;
\be
dP \propto  Q^3(\sigma) d\sigma \propto Q^3 \frac{d\sigma}{dQ} dQ
\label{anthropic}
.
\ee

Consider first the  case $C\ll H$. The regime
$\sigma \lsim  H$ is unviable because it gives $Q\sim 1 >Q\ma$.
Therefore we consider the regime  $\sigma\gg H$,
and assume that $\sigma\ma$ is big enough that its value does not matter.
(As $\sigma\ma$ is reduced from some such   value, it remains 
irrelevant to the following discussion until the 
 allowed rectangle ceases to intersect the curve making the model unviable.)
This is the version of the curvaton model considered by
 Garriga and Vilenkin \cite{gv}.
As we noted already it corresponds to $Q\simeq H/\sigma$
with $\oms$ very close to 1 and $\fnl=-5/4$. As $\sigma$ is reduced,
$Q$ rises to a maximum of order 1, but only the regime $Q<Q\ma$ is allowed.
Within this regime the 
  probability is
\be
dP\propto Q dQ
\label{gvest}
,
\ee
making  $Q= Q\sub{max}$  the most likely value. Taking
$Q\sub{max} = 10^{-4}$, the probability that $Q$ is at or below its observed
value is
 $1/25$. As these authors emphasize,
the  low probability
for the observed value need not be taken very seriously if only because
 $Q\ma$  is not at all well-defined.

Let us move on to the regime $C\gg H$.
In this case $Q(\sigma)$ has a maximum at $\sigma\sim C$, with the value
$ H/C$. Let us consider first the case that
 $\sigma\ma$ is big enough for its value not to matter.

 If $H/C  < Q\sub{max}$, the peak value is anthropically allowed,
and because of the $Q^3$ factor is in fact favoured;
 it had better agree with observation
if the anthropic argument is to work.  It corresponds to $\oms \sim 1$
and $\fnl\sim 1$. This is the case considered by Linde and Mukhanov
\cite{lm2}.
Now suppose instead that $H/C$ is bigger than $Q\sub{max}$, so that 
 a region around the peak is excluded.  
The  probabilities to the right and  to the left of
the peak are 
\bea
dP\sub{right} &\simeq&  \frac{H^3}{\sigma^3} d\sigma \\
dP\sub{left} &\simeq & \( \frac{H \sigma}{C^2} \)^3 d\sigma
\label{dpleft}
.
\eea
Integrating these expressions, the 
 the  relative probability  for being in the two regions are
\be
\frac{P\sub{left}}{P\sub{right}} \sim  \( \frac{Q\sub{max}C}{H}\)^2
< 1
.
\ee
The right-hand region is therefore preferred anthropically, leading
again to the estimate \eqreff{gvest}.

We have still to consider the case that $\sigma<\sigma\ma$ is a significant
constraint. As $\sigma\ma$ 
 moves down from a large value it will at some point exclude
the right hand part of the curve. Then
\eq{dpleft} applies which gives $dP \propto Q^3 dQ$. The probability 
that $Q$ is at or below the observed value is now only $1.6\times 10^{-3}$
(with $Q\ma = 10^{-4}$) which might be regarded as a catastrophe
for anthropic considerations. And  we are now in the regime of
strong non-gaussianity, which means that depending on parameters there
might be a violation of the present observational  bound on $\fnl$.

As $\sigma\ma$ moves further down, it will start to cut into the left hand
part of the curve. Eventually, 
 the peaked probability distribution
that we encountered in the previous paragraph is cut off at
 $\hat Q\ma \equiv Q(\sigma\ma) < Q\ma$ so that $\hat Q\ma$
becomes the preferred value, which had
better agree with observation if the anthropic argument is to
work. Again, one has to check that $\fnl$ is small enough.

This completes our discussion of the anthropic status of a  simple
version of curvaton model.
We see that the situation is rather  complicated.
 It would get still more complicated  if
we allowed  evolution of the curvaton 
(not to mention the possibility that the curvaton contribution is sub-dominant)
or if we considered a very non-flat prior probability for 
$\sigma$ such as might come  a departure from the 
probability distribution \eqreff{vgauss}.

\subsection{Comparison with a previous work}

To a considerable extent Sections \ref{s:bd}--\ref{s:an}
represent a development of \cite{lm2} (see also \cite{lm}).
Where they are comparable, the  results  are in broad agreement.

An expression essentially  equivalent to  \eq{supz} is derived
 in \cite{lm,lm2}.
To be precise, the expressions formally coincide because our 
 factor $e^{N\ts}$ is equal to their factor
$(H_0\mone/\lambda_0)$ defined in \cite{lm2}. But we
make a distinction between the number of $e$-folds of inflation (after
the observable Universe leaves the horizon) and the number of {\em relevant}
$e$-folds, because it is  the latter that should be identified with $N$.

The only other significant  difference is one of interpretation, concerning the
Bunch-Davies gaussian field $\sigma(\bfx)$  which they call the curvaton web.
One's view about the curvaton web depends on the interpretation of $\sigma$.
Within the observable Universe, the local value of  $\sigma$ will vary from 
place to place, and  its variation may be observable. The variation of 
$\sigma$ might be important  if, for instance,
one is considering 
 the  galaxy distribution in a relatively small region surrounding our
galaxy.  Indeed,  the average of
$\sigma$ within this  region (at horizon exit) may be significantly
different from its average within the whole observable Universe.
(Analogous considerations for the  axion dark matter density were pointed
out for instance in \cite{myaxion}.) With this interpretation of $\sigma$,
the steep spatial gradient of $\sigma(\bfx)$ 
evident in one of the simulations
may be an observable effect, as the authors remark.

On the other hand, the analysis of local effects within the observable
Universe is a rather tricky business even if one is not concerned with a
varying scalar field (but instead with say the local expansion rate).  
It may therefore 
 be useful to have a division
of labour, whereby  models of  the  early universe give the
correlators defined with the box size comfortably enclosing the whole
observable Universe. These then provide the starting point for the 
analysis of local effects, which can be 
done at a later and more sophisticated
stage of the research. Certainly that is the viewpoint usually taken
when the curvature perturbation is supposed to come from the inflaton,
and it has been the  viewpoint also of the present paper.
{}From this viewpoint, 
observation is sensitive to just one point on the curvaton web, whose
spatial gradient ceases to be physically significant.
The  gaussian probability distribution for $\sigma$ within the minimal box,
when inserted into \eqs{qexp}{fnlexp}
directly gives the cosmic uncertainty of the correlators, and with
the usual `cosmic variance' of the almost-gaussian CMB multipoles
this covers all possibilities for what will be observed even though it may
take some effort to work them out.

\section{Curvaton-type models after WMAP year three}

\label{s6}

If a curvaton-type contribution $\zetas$ dominates the curvature perturbation,
the tensor fraction is tiny. Then the WMAP year three results 
\cite{wmap3}
combined with
the SDSS galaxy survey give  $n-1\simeq -0.052^{+0.015}_{-0.018}$,
 and the result 
hardly changes if WMAP data are used alone or with several other relevant
data sets. 

If this measurement of a small negative spectral tilt holds up it has 
important consequences for curvaton-type models for the origin of the 
curvature perturbation. As was noticed in the early days of their exploration
\cite{liberated}, the most natural expectation for these models is that the
spectral tilt  $\ts$
 of the curvaton-type field $\sigma$ is practically zero.
 This is because, in contrast with the inflaton potential in a non-hybrid 
model of inflation, the potential of $\sigma$ does not 'know' about the 
end of inflation. The potential already has to be exceptionally flat just
to convert the vacuum fluctuation of $\sigma$ into a classical perturbation,
and in the absence of any reason to the contrary one might expect that the
departure from flatness will be too small to observe.

If this expectation is accepted, the spectral tilt predicted by a 
curvaton-type model is
\be
n-1 = - 2\epsilon_*
\dlabel{curvtilt0}
,
\ee
where $\epsilon$ is the flatness parameter of slow-roll inflation,
or more generally is  the parameter $\epsilon_H\equiv -\dot H/H^2$.
To reproduce the observed tilt we need a more or less scale-independent
value $\epsilon\simeq 0.025$

Among a suite of models considered in a recent survey \cite{al}, the
large-field models with $V\propto \phi^\alpha$ (chaotic inflation)
 give roughly the correct $\epsilon$.
The degree of tilt depends  on $N$, defined in this context as the number
of $e$-folds of inflation after the observable Universe leaves the horizon.
The best-motivated case is $\alpha=2$, because it may be obtained as an 
approximation to Natural Inflation.
Taking $N=50$, this gives
 $n-1=-0.020$ which is 
 a bit too small compared with the observed
value.
The multi-component  version of this  potential 
can help by reducing  $n-1$, but no investigation has been 
done to see how far one can go in that direction.

Increasing $\alpha$ increases $\epsilon$ by a factor $\alpha/2$,
so that $\alpha= 4$ or $6$ give a tilt agreeing with observation.
(If the inflaton perturbation is required to generate the curvature
perturbation such values of $\alpha$ are excluded, but that is not the case
here.) The problem with increasing $\alpha$ is that it lacks motivation,
either from string theory or from received wisdom about field theory
\cite{al}. If one accepts an increased $\alpha$ it may be more sensible to
regard $\alpha$ as a non-integer, providing an approximation to the 
potential over the relevant range of $\phi$.

What about the possibility of giving the curvaton-like field $\sigma$
a  significant
negative tilt $\ts$, say enough to allow $n-1\simeq \ts$? This requires
$\sigma$ to be on  a concave-downward part of its potential during 
inflation. Taking the view that the value of $\sigma$ is an initial
condition to be assigned at will (possibly with anthropic restrictions)
this need not be a problem. If instead the value has the stochastic
probability distribution \eqreff{vgauss}, 
there is some tension as we saw after \eq{pngb}
but one can still achieve the required small tilt with reasonable probability.
Only in the simplest version of the actual curvaton model does \eq{vgauss}
(with a quadratic or periodic potential) 
 demand negligible tilt $\ts$.
In general then, curvaton-type models can easily give the curvaton
perturbation a suitable negative tilt allowing $n-1\simeq \ts$.
The challenge in that case though, is 
to explain the actual value  $n-1 \simeq -0.05$ in a natural way.
As was noticed a long time ago \cite{treview}, a wide class of inflation
models generating the curvature perturbation from the inflaton do just that,
by making $n-1\sim -1/N$.

If curvaton-type models are rejected as the origin of the curvature
perturbation, that rejection is itself a powerful constraint on any
early-universe scenario involving scalar fields other than the inflaton.
Such fields must either be heavy during inflation, or else their
contribution to the curvature perturbation must be negligible.
The possible problem caused by a curvaton-type 
contribution being too big is analogous
to the moduli problem, and indeed might even be part of that problem
\cite{moroi,hkmt}.

\section{Conclusion}
\label{six}

The main result of Section \ref{s2} is the extension of \cite{bl}
to include strong spectral tilt, which is allowed if the curvaton
contribution gives a sub-dominant contribution to the curvature perturbation.

Section \ref{four} shows how to derive \eq{withlin} and its generalizations,
in the presence of both  spectral tilt and non-gaussianity. 
We note that the curvature perturbation generated by 
curvaton-type model long after inflation might have a very low ultra-violet
cutoff $\kmax$. This means that the scale of 
inhomogeneities in a matter component of the cosmic
fluid might not extend much below the scale $10^6\msun$ required for the 
formation of the first baryonic objects. On the other hand it might, in
which case 
 the  curvaton-type contribution to $\zeta$ might 
 be negligible on large cosmological scales, but big on sub-cosmological
scales even allowing primordial black hole formation.

On the technical side, we note that
the horizon-crossing formalism does not work
when spectral tilt and non-gaussianity
 are both to be included. Instead one should evaluate the field perturbations
at a fixed epoch,  which might as well be the one when $\kmax$ leaves the 
horizon. 

Section \ref{five} revisits  the actual curvaton model, taking into account
all of the possible effects that have been noticed. 
If the curvaton contribution to the curvature perturbation has strong positive
tilt, it can be negligible on cosmological scales but big enough to form
primordial black holes on smaller scales.
Recent 
discussions concerning cosmic uncertainty and the   anthropic status
of the curvaton models are extended.
This section also 
demonstrates  that the
 prediction $\fnl=-5/4$,  of the  simplest version of the 
curvaton model, can be obtained without recourse to the sudden-decay
approximation. Consequently, 
a detection $\fnl=-5/4$ would be a smoking gun for the simplest version of the 
curvaton model.
Finally, in Section \ref{s6} we looked at the status of  curvaton-type,
in the light of the recent measurement of negative spectral tilt for 
the curvature perturbation.

{\it Acknowledgments.}~ 
I thank  colleagues  K.\ Dimopoulos, E. Komatsu,
A.\ Linde,  K.\ Malik, S.\ .Mukhanov,   A.\ Starobinsky and  A.\
Vilenkin for valuable  communications.
 This work is supported
by PPARC grants  PPA/V/S/2003/00104,
PPA/G/O/2002/00098 and PPA/S/2002/00272 and EU grant 
MRTN-CT-2004-503369.

\end{document}